\documentclass{emulateapj}
\usepackage{subfigure}
\usepackage{url}
\usepackage{multirow}
\usepackage{color}
\usepackage{graphicx}
\usepackage{xspace}
\usepackage{natbib}
\usepackage{amsmath}
\usepackage{appendix}
\usepackage{multirow}
\usepackage{float}

\usepackage[figuresright]{rotating}

\DeclareMathOperator{\arccot}{arccot}
\newcommand{\water}{$\rm H_2O$} 
\newcommand{\methane}{$\rm CH_4$}
\newcommand{\ammonia}{$\rm NH_3$}
\newcommand{\cotwo}{$\rm CO_2$}

\shorttitle{2D Phase Curve Retrievals} \shortauthors{Feng et al.}

\begin{document}

\title{\uppercase{2D Retrieval Frameworks for Hot Jupiter Phase Curves}}

\author{\textsc{Y. Katherina Feng\altaffilmark{1,2}, Michael R. Line\altaffilmark{3}, Jonathan J. Fortney\altaffilmark{1}}}

\altaffiltext{1}{Department of Astronomy \& Astrophysics, 1156 High Street, University of California, Santa Cruz, CA 95064, USA.}
\altaffiltext{2}{NSF graduate research fellow}
\altaffiltext{3}{School of Earth \& Space Exploration, Arizona State University}


\begin{abstract}
    Spectroscopic phase curves provide unique access to the three-dimensional properties of transiting exoplanet atmospheres.  However, a modeling framework must be developed to deliver accurate inferences of atmospheric properties for these complex data sets.  Here, we develop an approach to retrieve temperature structures and molecular abundances from phase curve spectra at any orbital phase. In the context of a representative hot Jupiter with a large day-night temperature contrast, we examine the biases in typical one-dimensional (1D) retrievals as a function of orbital phase/geometry, compared to two-dimensional (2D) models that appropriately capture the disk-integrated phase geometry. We guide our intuition by applying our new framework on a simulated HST+Spitzer phase curve data set in which the ``truth'' is known, followed by an application to the spectroscopic phase curve of the canonical hot Jupiter, WASP-43b. We also demonstrate the retrieval framework on simulated JWST phase curve observations.  We apply our new geometric framework to a joint-fit of all spectroscopic phases, assuming longitudinal molecular abundance homogeneity, resulting in an a factor of 2 improvement in abundances precision when compared to individual phase constraints. With a 1D retrieval model on simulated HST+Spitzer data, we find strongly biased molecular abundances for \methane\ and \cotwo\ at most orbital phases. With 2D, the day and night profiles retrieved from WASP-43b remain consistent throughout the orbit. JWST retrievals show that a 2D model is strongly favored at all orbital phases. Based on our new 2D retrieval implementation, we provide recommendations on when 1D models are appropriate and when more complex phase geometries involving multiple TP profiles are required to obtain an unbiased view of tidally locked planetary atmospheres.
\end{abstract}

\section{Introduction}
Hot Jupiters have complex atmospheres; they are expected to be tidally-locked and experience large day-night temperature contrasts, along with significant  variations in abundances and cloud properties \citep{ParmentierCrossfield2018}. Phase curve observations of tidally locked exoplanets probe the longitudinal variations in temperature, composition, and cloud properties, acting as a powerful diagnostic of energy and chemical transport \citep[e.g.,][]{agundez2012,komacek2017,drummond2018,steinrueck2019}.  Furthermore, precision abundance ratios can potentially be tied back to models of planet formation \citep[e.g.,][]{Oberg2011,Madhu14,Espinoza17}. With the promise of better precision on the horizon from future observatories such as the \emph{James Webb Space Telescope}, we will be able to deepen the level of our characterization of exoplanet atmospheres. It is critical to our understanding of these worlds to assess the accuracy with which this information can be constrained by leveraging the synergies between observations and modeling efforts. 

Atmospheric retrievals have emerged as a powerful tool for determining atmospheric properties such as molecular/elemental abundances, cloud properties, and thermal structures from exoplanet spectra \citep{madhuseager2009,line2012,lee2012,benneke2015,madhusudhan2018}. Inverse modeling is driven by the data set, wavelength coverage, and observation uncertainty; just as much, retrieval-based atmospheric inference is highly model dependent, a realization that has recently received well-deserved attention. As more inverse models are developed and data diversity continues to increase, we face a growing suite of choices regarding radiative transfer, chemistry, and aerosol treatment. \citep{changeat2019,molliere2019,mai2019,iyer2019,barstow2020}. The specifics are ever evolving as it can be complex to pinpoint what a model may be lacking within the context of a specific data set. 

A challenging aspect of retrievals is maintaining a computationally efficient forward model within common Bayesian frameworks while balancing adequately sophisticated implementation of the necessary atmospheric physics. Given the disk-integrated nature of the observed spectra, retrieval models have typically assumed 1D treatment of the temperature-pressure (TP) profile and chemistry. Yet, for instance in the case of a planet observed at quadrature, where half the dayside and half the nightside are visible, the hemispherically averaged spectrum would include contribution from contrasting hot and cool temperature-pressure (TP) profiles. Consequently, we used this case in our previous work \citep{Feng2016} to demonstrate that a 1D retrieval model assumption affects atmospheric inference and can introduce unwanted biases. The significance of the impact depends on the type of data set and temperature contrast between the day and the night. This work found that methane is mischaracterized for simulated \textit{Hubble Space Telescope} Wide Field Camera 3 (WFC3) and \textit{Spitzer Space Telescope} Infrared Array Camera (IRAC) data (hereafter HST+Spitzer) - biased to a precise but inaccurate posterior distribution. Furthermore, for simulated JWST data, even water is mischaracterized.  

In addition to \citet{Feng2016}, numerous other works have explored the impacts of the 1D treatment for inherently-3D exoplanet atmospheres.  \citet{lineparmentier2016} demonstrated how nonuniform terminatory cloud cover can mimic high mean molecular weight atmospheres in transmission spectra observations. \citet{Blecic2017} used 3D general circulation model (GCM) outputs to generate synthetic emission spectra at secondary eclipse on which to test a 1D retrieval, finding that the 1D TP profile resembles the arithmetic average over the profiles within the 3D model. These results were dependent on the data quality, from the wavelength coverage to instrument resolution. \citet{caldas2019} identified biases in interpreting transmission spectra associated with the day-night temperature gradient through the limb of the atmosphere. Specifically, the 1D retrieval models perfectly fitted 3D generated spectra, but resulted in biased retrieved abundances. {As pointed out in \citet{macdonald2020} and \citet{pluriel2020}, the terminator probes not only differences between day and night temperatures, but also resulting divergent chemical composition. By examining the impact of inhomogeneous terminator composition on transmission spectra retrievals, \citet{macdonald2020} and \citet{pluriel2020} revealed that both retrieved temperatures and abundances suffer from substantial biases when assuming a 1D forward model.} 

One promising avenue for elucidating the 3D structure of an exoplanet's atmosphere is the spectroscopic phase curve. With a different hemispheric average observed at each phase, we use phase curves to investigate energy transport, atmospheric dynamics, chemistry, and cloud distribution and composition. 

Typically, sophisticated 3D general circulation models have been the preferred approach for interpreting spectrophotometric phase curve observations \citep[e.g.,][]{showman2009,kataria2015, mendonca18mixing}. \citet{Stevenson2014,Stevenson2017} provide the first spectrophotometric phase curve data set of a hot Jupiter \citep[WASP-43b,][]{Hellier2011} of which at the time was interpreted with the simplistic 1D model. This likely has resulted in strong abundance biases, as shown in \citet{Feng2016}. Retrievals that can accommodate the inherent 3D nature of phase data are relatively new due to the challenge of adding the necessary complexity in a computationally efficient manner as well as higher fidelity data (as for WASP-43b). 

Recently, \citet{Irwin2020} adapted {the optimal estimation variant of the NEMESIS retrieval code \citep{irwin2008}} to perform ``2.5-D'' retrievals on the spectroscopic phase curve of WASP-43b. By using a parameterized prescription of assigning temperature and composition as a function of longitude and latitude, \citet{Irwin2020} are able to retrieve thermal structures for WASP-43b that are more consistent with GCM predictions than the simplified 1D models. However, the optimal estimation method is limited in its ability to provide a wide enough sampling of parameter space or a means of model comparison based on data quality. {In the coming years, we anticipate more retrieval models to account for the multi-dimensional nature of atmospheres. \citet{changeat2020}, for example, present a semi-analytical model for phase-dependent emission based on the projected 2D disk of a planet. Their approach also enables contribution from the day-night terminator. \citet{changeat2020} do not perform retrievals on simulated or observed phase curve data, although the model is slated to be applied within the TauREx 3 retrieval framework \citep{taurex} in the future.}

While state-of-the-art phase curves obtained with HST+Spitzer only exist for a handful of exoplanets \citep[e.g.,][]{maxted2013,Stevenson2017,kreidberg2018,arcangeli2019}, JWST will enhance this technique and measure phase curves across longer wavelength ranges \citep{bean2018}. {\citet{beichman2014} and \citet{Irwin2020} identify JWST as an important avenue to explore in the context of spectroscopic phase curves.} \citet{venot2020} simulated JWST MIRI ($5-12\, \mu$m) observations of the phase curve of WASP-43b; while thorough, their retrieval study uses the typical 1D approach. Potential pitfalls resulting from a 1D approach need clarifying. Recently, \citet{Taylor2020} consider the possibility of identifying non-uniform thermal structure from emission spectra of WASP-43b given different observing modes of JWST NIRSpec. \citet{Taylor2020} also find evidence of biased molecule abundance estimates described in \citet{Feng2016}, detailing the dependence on model choice and wavelength range. 

As such, we seize a unique opportunity in this study to complement previous work by incorporating phase geometry within a Bayesian retrieval framework such that we can robustly explore the following questions: What inferences remain consistent as a function of phase for a planet with 2 contrasting TP profiles? What will we gain when we use JWST? Is there any advantage in leveraging the full set of phase curve data together? 

We build on previous work by \citet{Feng2016} to systematically explore biases resulting from retrieval assumptions by introducing additional model complexity. {We see this work as an integral addition to the growing body of works dedicated to multi-dimensional retrievals. In particular, it is parallel in effort to the ``2.5-D'' retrieval framework developed by \citet{Irwin2020}. Instead of a parametric prescription to determine abundances and temperature as a function of longitude and latitude, we utilize location-invariant abundances to better focus on the impact of thermal inhomogeneity in phase curve retrievals. The parameterization of TP profiles ensures consistency with \citet{Feng2016} methods and a more direct comparison with our previous work. With our nested sampling retrieval, we are able to marginalize and illustrate the posterior distribution of parameters as a function of phase. A Bayesian nested sampling framework will also provide metrics for model comparison, an essential component in evaluating the effectiveness of more complicated models.}

In this paper, we present a new framework that uses spherical trigonometry to properly model the phase geometry within the CHIMERA retrieval suite \citep{Line2013}. Section \ref{sec:methods} describes the relevant adaptations and our investigation setup. We simulated HST+Spitzer phase curves to anchor our understanding of thermal inhomogeneity based on the 1TP and 2TP models. We then retrieve on \citet{Stevenson2017} WASP-43b data as well as simulated JWST phase curves of a model planet. Section \ref{sec:results} compiles these results. As in \citet{Feng2016} and \citet{Taylor2020}, we focus on model comparison in the this study; we provide a guide as to which phases need the appropriate 2TP modeling of the large day-night temperature contrast to accurately interpret the atmosphere. We conclude with Section \ref{sec:discussion} through a discussion of our findings and future work.

\section{Methodology}
\label{sec:methods}
In \citet{Feng2016}, we tackled thermal inhomogeneity with a simplified experiment: A fully symmetric scenario where half the emitting area is attributed to a hotter TP profile, while the other half is from a colder profile, permitting a simple averaging of the resultant spectra. Such a scenario would be applicable to the quadrature phases, assuming the day and night side can each be well-represented by a single thermal profile, as well as a planet ``checkered'' with hot and cold patches. For non-symmetric cases, we need a more sophisticated geometry to account for the differing contributions as well as the appropriate limb darkening. In the subsections that follow, we provide an overview of the radiative transfer and retrieval analysis. Next, we describe the modifications to the model used in \cite{Feng2016} to generate spectra at arbitrary orbital phases. We also detail our validation and lay out the investigation plan. 

\subsection{Overview of modeling tools}
 The core radiative transfer routines remain identical to those in \citet{Feng2016}. \citet{Line2013} provide a detailed description. Given a TP profile and vertically uniform gas mixing ratios, we solve for the outgoing thermal radiation in a cloud-free, plane parallel atmosphere. We retrieve for \water, \methane, CO, \cotwo, and \ammonia, and assume solar composition H$_2$/He as the background filler. As in \citet{Feng2016}, opacities for these gases are drawn from the database described in \citet{Freedman2014} and \citet[][Table 2]{lupu2014}. 

As in our previous work, we consider the distinction between 1TP and 2TP. We focus on a large temperature contrast case (80\%) to see what effects may be dominant across phase. {Conceptually, we think of the contrast factor can as} $1 - \frac{T_{\rm TOA, c}}{T_{\rm TOA, h}}$ \citep{Feng2016}.  The second term is a ratio between the temperature at the top of the atmosphere (TOA) from the cool profile (c) and the TOA temperature from the hot profile (h). Table \ref{tab:params} {lists fixed system properties and the parameters we retrieve for in the 2TP model, including the parameterization of the contrast.} We calculate the TP profile using the approach from \citet{Parmentier14} (see also Equations 13 and 14 in \citet{Line2013}). Each profile is defined by five parameters: two visible-to-infrared mean opacity ratios ($\log \gamma_1$ and $\log \gamma_2$), the partitioning between the two visible streams ($\alpha$), the infrared opacity ($\log \kappa_{\rm IR}$), and the fraction of absorbed incident flux ($\beta_{\rm TP}$). We specify $\beta_{\rm day}$ for the day profile and $\beta_{\rm night}$ for the night profile while letting the two profiles share the other parameters. For values $\beta_{\rm day} = 1$ and $\beta_{\rm night}=0.2$, we establish the contrast of 80\% between the day and night sides.  

In other words, {the 2TP model retrieves for 11 parameters (5 molecules + 6 TP parameters) while the 1TP model retrieves for 10 (5 molecules  + 5 TP parameters).} All scenarios assume an internal temperature, $T_{\rm int}=200$K (although higher values can be expected in hot Jupiters; see \citet{thorngren2019}). We assume constant-with-altitude and constant-with-longitude \citep[e.g.,][]{coopershowman2006,mendonca18mixing} mixing ratios loosely consistent with thermochemical equilibrium and solar composition abundances. We adopt WASP-43b planetary properties \citep{Hellier2011} in our model (see also Table \ref{tab:params}). 

\begin{figure}[!tbp]
  \centering
  \includegraphics[width=0.4\textwidth]{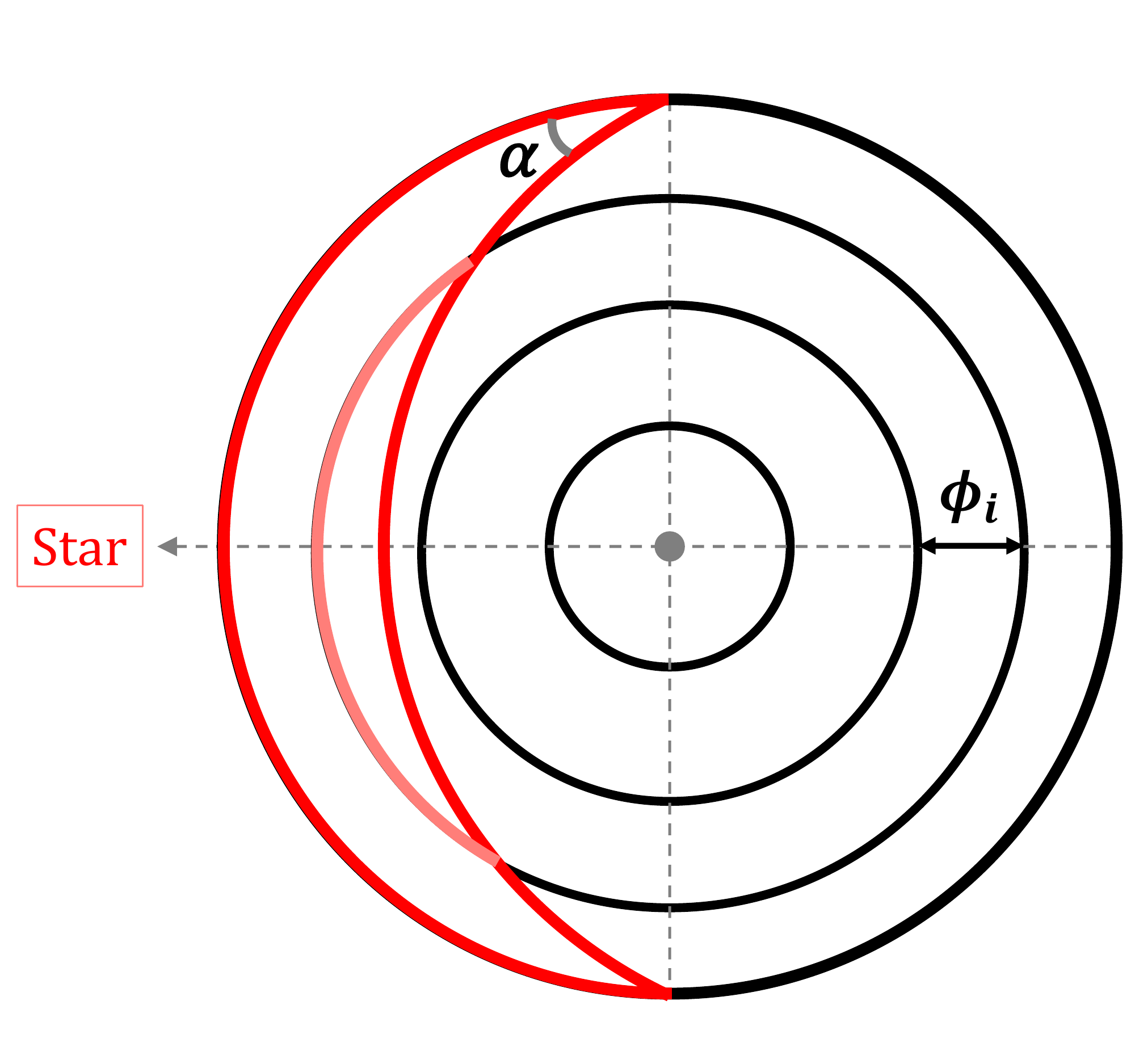}
  \caption{Diagram of hemisphere visible to the observer at phase angle $\alpha$. The visible hemisphere is divided into annuli based on Gaussian quadrature angles, which are used to define the annulus width $\phi_i$. The ``dayside'' region (in red) intersects the annuli at different points. By determining the areas of these segments within each annulus, we can determine the fractional contribution of ``day'' and ``night'' for the annulus, as described more in Appendix \ref{sec:geometry}. Within the ``2TP'' approach, all annuli areas encompassed by red will be assigned a ``dayside'' TP, and in black, a ``nightside'' TP.  }
  \label{fig:geo_front}
\end{figure}

Typical radiative transfer uses an ``N-point'' (4 in \citet{Feng2016}) Gaussian quadrature to compute the TOA outgoing fluxes whereby the observed disk can be divided up into concentric ``annuli'' (Figure \ref{fig:geo_front}) with intensities computed independently at each annulus (given the quadrature $\mu$). An example crescent phase in Figure \ref{fig:geo_front} shows how each annulus contains different contributions of ``dayside`` and ``nightside'' regions (and properties thereof). In this work, we adopt the same Gaussian quadrature scheme but have to apply geometric corrections to account for the varying viewing geometry as a function of phase (e.g., for uneven day-night temperature variations). Rather than ``pixelating'' the planet \citep[e.g.,][]{fortney2006,showman2009,cahoy2010}, we divide it up in annuli (a natural radiative transfer coordinate system) whereby we can assign individual atmospheric properties (temperature, composition, etc.) that will dictate the upwelling intensity beam. Summing over these beams will produce the appropriate disk-integrated flux accounting for atmospheric inhomogeneity. Appendix \ref{sec:geometry} describes in detail the adjustments needed to accommodate arbitrary phase angles within the Gaussian quadrature/concentric annuli radiative transfer framework. We will refer to this updated model as 2TP-Crescent. 

Table \ref{tab:phases} lists the orbit fraction and phase angle we consider, consistent with the phase curve data set presented in \citet{Stevenson2014,Stevenson2017}. Hereafter, we will refer to orbital phases by a corresponding number as listed in Table \ref{tab:phases}. The new geometric implementation also allows for the retrieval of day-side hot spot properties {in similar spirit of eclipse mapping} (modeled as inner annuli with higher temperature than remaining annuli, Figure \ref{fig:geo_front}), a thorough exploration of which is beyond the scope of our current study. 

{We pair our modified forward model with {\tt pymultinest} \citep{PMN}, the python implementation of the {\tt multinest} algorithm \citep{MN,feroz2009,feroz2019}, to perform Bayesian parameter estimation and model selection, following standard guidelines \citep[e.g.,][]{Trotta2008,Cornish07,trotta2017}. Each retrieval used 2000 {\tt multinest} live points.}

\def\arraystretch{1.5}
\begin{deluxetable}{lll}
\tablecaption{Model parameter values and priors \label{params}}
\tablewidth{0.35\textwidth}
\tabletypesize{\scriptsize}
\tablehead{Parameter & Value & {Prior}}
\startdata
\label{tab:params}
$R_p$ ($R_{\rm J}$)&  0.93& --\\
$R_*$ ($R_{\odot}$)&0.598 &--\\
$T_*$ (K) & 4400 & --\\
$a$ (AU)  &0.0142 & --\\
$T_{\rm int}$ (K) &200 &-- \\  
$\log(g)$ (cm s$^{-2}$)  &{3.672} & --\\  
$\log f_{\rm H_2O}$ & -3.37 & [-12,0] \\
$\log f_{\rm CH_4}$ & -9 & [-12,0]\\
$\log f_{\rm CO}$ &-3.7& [-12,0] \\
$\log f_{\rm CO_2}$ & -9& [-12,0]\\
$\log f_{\rm NH_3}$  &-9& [-12,0]\\ 
 $\log \gamma_1$ & -1 &[-3,2] \\
  $\log \gamma_2$ & -1 &[-3,2] \\
  $\log \kappa_{\rm IR}$ & -1 &[-3,0] \\
   $\alpha$ & 0.5 & [0,1]\\
   $\beta_{\rm day}$ & 1 & [0,2]\\
    $\beta_{\rm night}$ & 0.2 & [0,2]\\
    f${_{\rm day}} \footnote{The parameter f$_{\rm day}$ represents the fractional area of the visible ``dayside''. It varies as a function of phase according to Equation \ref{eq:illum}.}$ & see eq. \ref{eq:illum} & [0,1]
\enddata
\tablecomments{Nominal system and TP shape parameters used to generate our synthetic spectra. Stellar and planetary parameters are based on the WASP-43 system \citep{Hellier2011}. For definitions of the TP parameters, see \citet{Line2013}. Solar proportion Hydrogen and Helium are assumed to make up the remaining gas abundance {(He/H$_2$ = 0.176471). We focus on C-N-O-bearing molecules. Retrieved parameters include prior ranges. Priors are uniform or log-uniform.}}
\end{deluxetable}

\def\arraystretch{1.5}
\begin{deluxetable}{cll}
\tablecaption{Reference for phase angles}
\tablewidth{0.35\textwidth}
\tabletypesize{\scriptsize}
\tablehead{Phase & Angle ($^\circ$) & Fraction}
\startdata
\label{tab:phases}
0  &  22.5 & 0.0625    \\
1  &  45.0 & 0.125   \\
2  &  67.5 &  0.1875    \\
3  &  90.0 &  0.25 (quadrature)   \\
4  &  112.5 &  0.3125    \\
5  &  135.0 &  0.375    \\
6  &  157.5 &  0.4375    \\
7  &  180.0 &  0.5 (secondary eclipse)   \\
8  &  202.5 &  0.5625    \\
9  &  225.0 &  0.625    \\
10  &  247.5 &  0.6875    \\
11  &  270.0 &  0.75 (quadrature)   \\
12  &  292.5 &  0.8125    \\
13  &  315.0 &  0.875   \\
14  &  337.5 &  0.9375    
\enddata
\tablecomments{We assume the full orbit ($360^{\circ}$) is divided into 15 phases. Phase 0 is right after transit; phase 7 is secondary eclipse; and phase 14 is right before transit.  Orbital fraction is the phase value between 0 and 1. The phase angle (between the sub-observer point and sub-stellar point) is defined from transit; see Figure \ref{fig:annuli}.  }
\end{deluxetable}


\subsection{Investigation set-up}
\label{subsec:investigate}
Within our new framework, we explore three phase curve observational setups under four different retrieval model assumptions. The three observational scenarios are:

\begin{itemize}
    \item Simulated (of which we know the ``truth'' values) HST+Spitzer observations based on the \citet{Stevenson2017} WASP-43b data set
    \item The actual \citet{Stevenson2017} WASP-43b HST+Spitzer phase curve data set
    \item Simulated JWST phase curve observations of WASP-43b
\end{itemize}

For the simulated data, we only consider phases between transit and secondary eclipse due to symmetry over the orbit (i.e., we do not assume hot spot offsets or other asymmetries). {The orbit is assumed to be circular, with $90^\circ$ inclination, and the planet is assumed to have zero obliquity.} We assume cloud-free atmospheres in order to not confuse any degeneracy arising between the geometric/model assumptions and basic atmospheric properties, like abundances and thermal profiles.  Future work looking into the nature of inhomogeneous clouds is most certainly a next step. 

The synthetic HST+Spitzer uncertainties are pulled phase-by-phase from the \citet{Stevenson2017} WASP-43b data set. For the simulated JWST data, we use the same data setup as described in \citet{Feng2016} (based on the \citet{greene2016} noise model), which assumes a single transit each in NIRISS, NIRCam, and MIRI LRS, covering 1-10 $\mu$m at a resolution ($R$) of 100. For the synthetic data, we do not apply random noise to the instrument-resolution points. As in past works \citep{Feng2018,krissansen2018,mai2019,changeat2019}, we opt not to randomize the simulated data points as to mitigate random bias due to outlier noise instance draws.

\def\arraystretch{1.5}
\begin{deluxetable}{lcccc}
\tablecaption{Data sets and relevant model scenarios}
\tablewidth{0.45\textwidth}
\tabletypesize{\scriptsize}
\tablehead{Scenario & HST\ $+$ & WASP-43b & JWST & Joint\\
                     & Spitzer &      &      & Phases }
\startdata
1TP & $\times$ & $\times$ & $\times$ & --\\
2TP-Crescent & $\times$ & $\times$  & $\times$ & $\times$ \\
2TP-Free & -- & $\times$ & -- & -- \\
2TP-Fixed & -- & -- & -- & $\times$ \\
\enddata
\label{tab:scenarios}
\end{deluxetable}

Table \ref{tab:scenarios} lists the data sets considered in our study along with corresponding model scenarios whose results are presented in Section \ref{sec:results}. In total, we examine four retrieval model assumptions:

\begin{itemize}
    \item 1TP: assumes a single TP profile and and one set of gas mixing ratios regardless of the observed phase, like in \citet{Feng2016} and applied in \citet{Stevenson2014,Stevenson2017}.
    \item 2TP-Crescent: retrieves a hot day profile and cool night profile. Hot and cool fluxes are distributed according to phase and combined with annuli (Appendix \ref{sec:geometry}), such that any limb darkening differences between hot and cold are included. 
    \item 2TP-Fixed: retrieves two profiles. Hot and cool fluxes are combined via linear combination using $F_{\rm total} = F_{\rm hot}*f_{\rm day} + (1-f_{\rm day})*F_{\rm cool}$.  Equation \ref{eq:illum} determines phase-dependent day-side contribution $f_{\rm day}$
    \item 2TP-Free: retrieves two profiles. Fluxes calculated in the same way as 2TP-Fixed, but the retrieval treats the dayside area fraction $f_{\rm day}$ as an additional free parameter \citep[as in][]{Taylor2020}. 
\end{itemize}

We first retrieve on each ``phase'' independently (phase-by-phase). In this situation, the ``day'' and ``night'' TP profiles are allowed to vary from phase to phase. We then use the 2TP-Crescent and 2TP-Fixed schemes to perform a joint retrieval on all phases simultaneously, assuming the same day and night TP profiles at each phase. {The likelihood used in the retrieval is the sum of individual likelihoods from each phase, $\log \mathcal{L}_i$, such that $\log \mathcal{L}_{\rm total} = \sum_i \mathcal{L}_i$, where} 

\begin{equation}
    \log \mathcal{L}_i = -\frac{1}{2} \sum_j \frac{(y_{i,j} - y_{{\rm mod}, i,j})^2}{\sigma_{i,j}^2}.
    \label{eq:jointL}
\end{equation}

{In Equation \ref{eq:jointL}, $y_{i,j}$ are the measured data points with uncertainties $\sigma_{i,j}$ for a given phase $\alpha_i$. Our forward model uses $\alpha_i$ to select the appropriate pre-computed dayside fractions (see Appendix \ref{sec:geometry}) for that phase angle and calculate model data points, $y_{{\rm mod},i,j}$.} The joint retrieval investigation aims to determine if improved abundance constraints are achievable if we assume a priori that the ``day'' and ``night'' profiles remain the same throughout the orbit.

\section{Results}
\label{sec:results}
Sections \ref{subsec:hst-results} through \ref{subsec:simretrieve} provide detailed retrieval results for each of the data sets we explored. We show the posteriors of the abundances as a function of phase, retrieved pressure-temperature profiles at several points in the orbit, and the spectral fits to the data for select phases. {For reference, Figure \ref{fig:fullpost} is an example of the full posterior distribution from one of our retrieval runs.}

The first data set is a simulated HST+Spitzer phase curve, where all input parameters are known.  Next, we perform a similar analysis on real WASP-43 HST+Spitzer phase curve data and examine the differences. We expand next to a simulated JWST data set with a higher signal-to-noise and wider wavelength range. Finally, we explore the concept of ``joint retrieval'' where we use the full suite of phase curve data to seek tightened error bars on atmospheric quantities.

\begin{figure*}[t]
  \centering
  \includegraphics[width=0.95\textwidth]{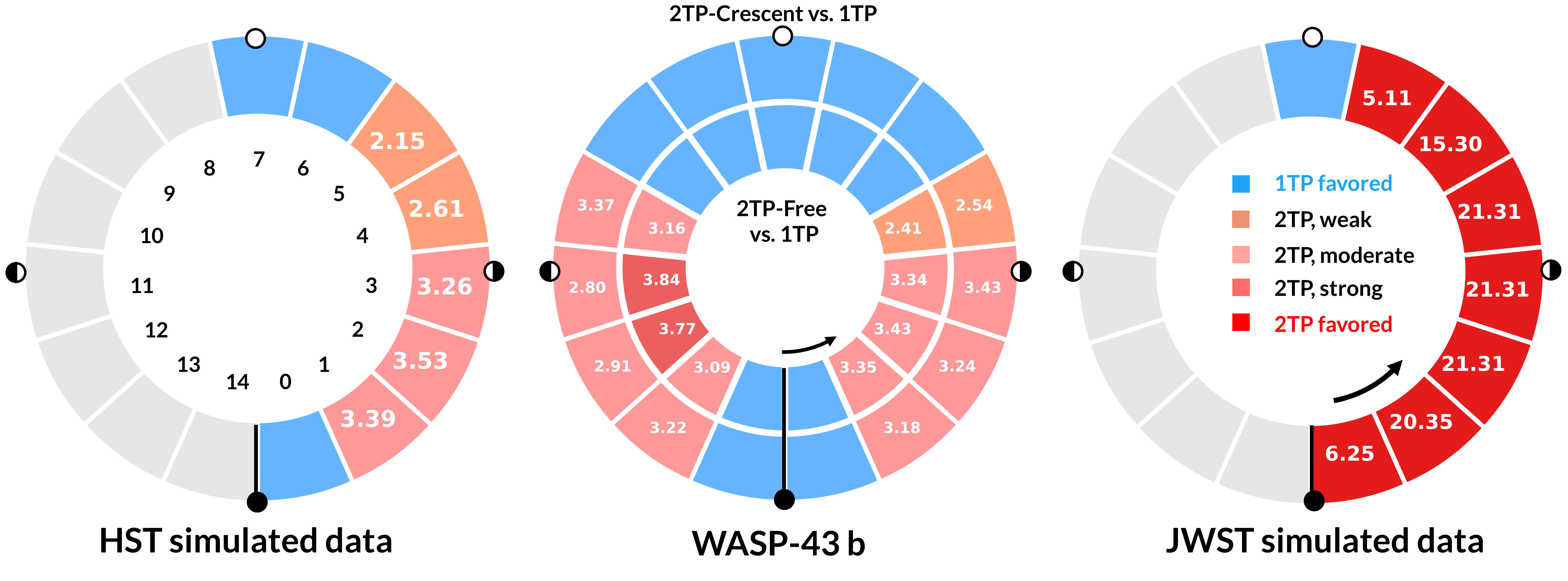}
  \caption{Model comparisons as a function of phase.  See text for how detection significances are computed. The significance values represented here follow: $\sigma_{\rm 2TP} \leq 2\sigma$: inconclusive (blue), $2\sigma <\sigma_{\rm 2TP} < 2.7\sigma$: {weak evidence (light orange)}, $2.7\sigma \leq \sigma_{\rm 2TP} < 3.6\sigma$: moderate evidence (light pink), $3.6\sigma \leq \sigma_{\rm 2TP} < 5\sigma$: strong evidence (deep pink), and $\sigma_{\rm 2TP} \geq 5\sigma$: significant evidence (red). Left: Detection significance of the 2TP-Crescent model compared to the 1TP model for the simulated HST+Spitzer data. Due to symmetry, we only simulated half the orbit. Phase are labeled by their numbers (Table \ref{tab:phases}).  Middle: Detection significance for observed WASP-43b HST+Spitzer data. Outer ring compares 2TP-Crescent to the 1TP model. Inner ring compares the 2TP-Free and 1TP models. Full orbit is considered. Right: 2TP-Crescent vs 1TP comparison on simulated JWST data. Due to symmetry, only half the orbit is considered. }
  \label{fig:sigmaCompare}
\end{figure*}

The focus of our study is on the difference between 1TP and 2TP models as a function of phase and determining the phases at which we are justified in employing a more complex 2D model. We thus synthesize our different scenarios by presenting an overview comparing the use of a homogeneous 1D model and a more complex model (2TP-Crescent and 2TP-Free specifically) on phase-resolved spectra. Figure \ref{fig:sigmaCompare} summarizes the justification of the 2TP model over the 1TP model as a function of phase for each of our cases. We will return to the figure throughout the paper. Following \citet{Trotta2008} and \cite{GordonTrotta2007}, we convert the Bayes factor into detection significance of the 2TP model. For each set of data we consider, we illustrate the degree to which it is justified to use the more complex 2TP model (Crescent or Free) to interpret the data. 

For example, for HST simulated data, the leftmost panel of Figure \ref{fig:sigmaCompare}, phases 1-3 moderately favor the addition of the night profile (indicated by light pink color). {Phases 4 and 5 (orange color) weakly favor the inclusion of a second profile. For phases 6 and 7, the closest to secondary eclipse, there is insufficient evidence} to suggest the 2TP model should be used for the data (indicated by blue color). However, as we will see in the posteriors of the abundances for the simulated HST+Spitzer data, {the phases that favor the second profile} reveal biased methane constraints when we use the 1TP model.

\subsection{Control Case: HST and Spitzer Synthetic Data}
\label{subsec:hst-results}

\begin{figure*}[b!]
  \centering
  \includegraphics[width=0.9\textwidth]{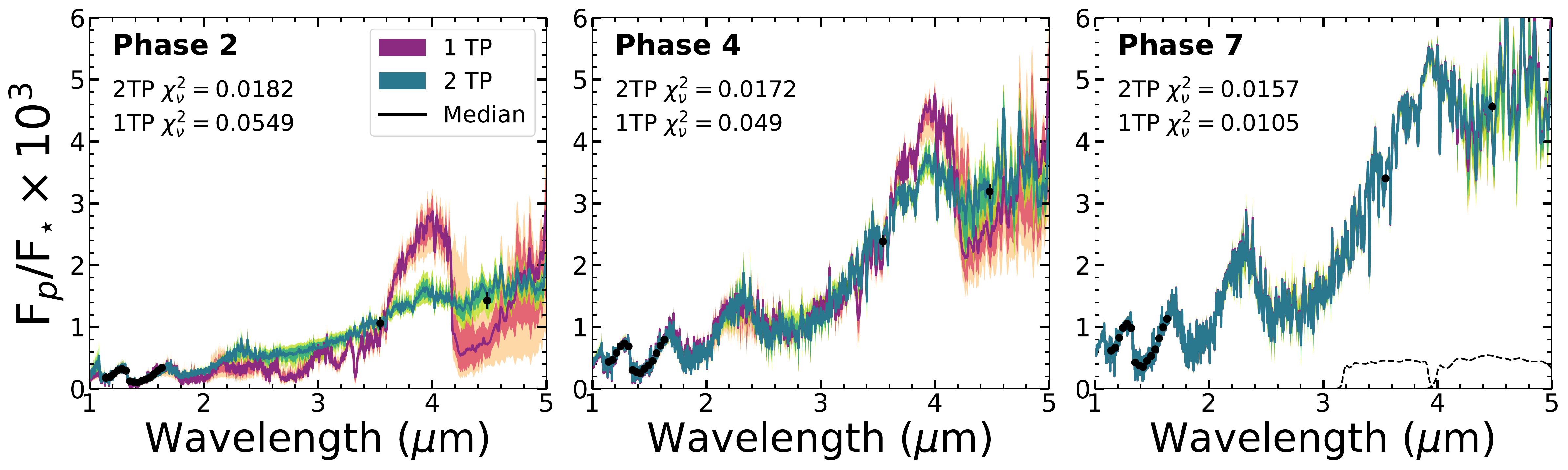}
  \caption{Simulated HST+Spitzer data and resultant representative fits drawn from the posterior for phases 2 (just before first quarter), 4 (just after first quarter) and 7 (secondary eclipse). We include the Spitzer 3.6$\mu$m and 4.5$\mu$m filter profiles in the phase 7 panel. 1TP spectra are in magenta while 2TP-Crescent spectra are in green. For each set of model spectra, we plot the median (solid line), $1\sigma$, and $2\sigma$ contour. We include corresponding $\chi^2_\nu$ values for the 1TP and 2TP(-Crescent) models, which can be small because random noise is not included. The spectra from the two models differ the most at phases close to transit; they become more similar as the phases advance to secondary eclipse, where they overlap. The biased \methane\ and \cotwo\ abundances result in more spectral contrast between 3 and 5$\mu$m for the 1TP profile scenario. With such distinct spectra at phases showing more night side emission, data filling the gaps between HST and Spitzer observations would be able to differentiate between these two models. }
  \label{fig:hst_spectra}
\end{figure*}

The results from our synthetic data set of HST+Spitzer observations serve as a guide for our intuition, allowing us to identify trends and biases before considering measured spectroscopic phase curves. Posterior-representative spectra are shown in Figure \ref{fig:hst_spectra} for both the 1TP and 2TP-Crescent scenarios. The 2TP profile properly fits the data but the 1TP case struggles for phases between transit and first quarter (0 - 2). Given the sparse data coverage with HST and Spitzer, model differences are most noticeable in unmeasured spectral regions, with resultant 1TP spectra presenting deeper absorption features owing to the steeper temperature gradients retrieved in the 1TP model.  At phases closer to secondary eclipse, the discrepancy wanes as the dayside TP/spectra more prominently represent the total. 

\begin{figure*}[!t]
  \centering
  \includegraphics[width=0.9\textwidth]{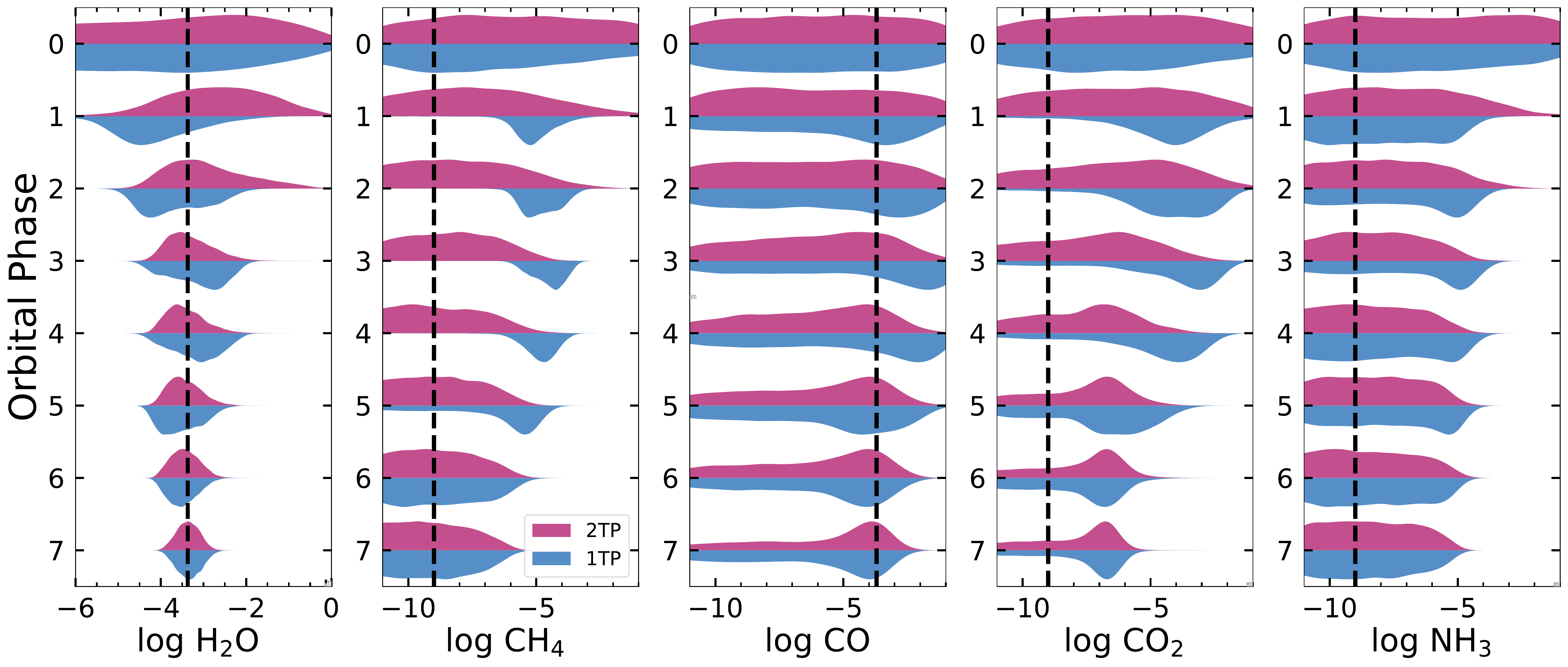}
  \caption{Abundance vs. phase results from HST+Spitzer simulated data for \water, \methane, CO, \cotwo, \ammonia\ for the 1TP model (blue) and the 2TP-Crescent model (dark pink). For each panel, we plot {the kernel density estimation of} the marginalized posterior probability distribution for the log of the molecule's mixing ratio as a function of orbital phase. {The distributions are set to show the same total height at each phase and thus do not show the relative probability.} For simulated data, we only consider half an orbit (transit to secondary eclipse), or eight orbital steps. For each molecule, we indicate the input abundance value with the vertical dashed line. This simulated data set is only able to accurately constrain \water\ abundance; both 1TP and 2TP-Crescent models provide consistent posteriors for \water. The other molecules have only upper limit estimates with the 2TP-Crescent model. For most of the phases, the 1TP model produces biased \methane\ abundances (constrained at values orders of magnitude above the input). \cotwo\ is biased toward higher values under the 1TP model for half the phases, {while under the 2TP model we see biased distributions for phases 5-7.}}
  \label{fig:hst_abunds}
\end{figure*}

Figure \ref{fig:hst_abunds} presents the 1TP and 2TP posteriors of the molecular abundances as a function of phase for the simulated HST+Spitzer data. Phase 0, just after transit, exhibits no constraining power for the mixing ratio of any molecule under either model due to the low feature signal-to-noise. The posteriors are identical for phase 7 (secondary eclipse), as expected, given that there is no visible night-side flux.

\water\ and CO are the only two molecules with high enough abundances in the input model such that we should expect detection. We find no significant bias in the \water\ abundance when using the overly simplistic 1D TP profile.  The ``truth'' falls within the $1\sigma$ range of the retrieved distribution at most phases ($2\sigma$, at worst). Phases for which there is more viewable ``dayside'' (hotter), the constraints are more precise {($\sim 0.3-0.5$ dex)}, and gradually decline towards the cooler phases (phase 0), simply due to the reduced feature signal-to-noise. {Table \ref{tab:h2ovals} in Appendix \ref{sec:supp} shows the phase-by-phase values for the \water\ abundance's under the 1TP and 2TP-Crescent models.} CO is largely unconstrained at all phases in both models as only the Spitzer 4.5$\mu$m photometric point is sensitive to this molecule . 

The original input mixing ratios for \methane, \cotwo, and \ammonia\ are all well below typical detectable amounts ($10^{-9}$), such that only upper limits would be anticipated. As such, biases become obvious when the retrieved distributions for some of these molecules are tightly constrained at elevated abundances. For example, we find that in Figure \ref{fig:hst_abunds} between phases 1 and 5, the 1TP model retrieves a tight constraint on \methane\ a few orders of magnitude higher than the input (median of $\sim 3.5\times10^{-5}$ vs. $10^{-9}$), as seen in our 2016 paper. However, using the correct 2TP-Crescent model (which was used to generate the simulated data), we retrieve only an upper limit at all phases, as expected. \ammonia\ shows similar behavior as \methane:  The 1TP model appears to produce more of a constraint at higher values (while still an upper limit) than the 2TP-Crescent model, as seen in phases 2 - 5 in Figure \ref{fig:hst_abunds}. This suggests that the 1TP model results in an \ammonia\ bias as well, though not as extreme as in the case for \methane.

CO and \cotwo\ show similar trends in Figure \ref{fig:hst_abunds} due to their overlapping spectral features over single 4.5$\mu$m Spitzer point \citep{Line2016}. In fact, \cotwo\ presents more bias when using the incorrect 1TP model. This apparent bias appears even at full phase (phase 7) in the correct 2TP-Crescent model because a large sample of models accumulate at high CO and \cotwo\ abundances simply due to the overlapping degeneracy.

\begin{figure}[t!]
  \centering
  \includegraphics[width=0.45\textwidth]{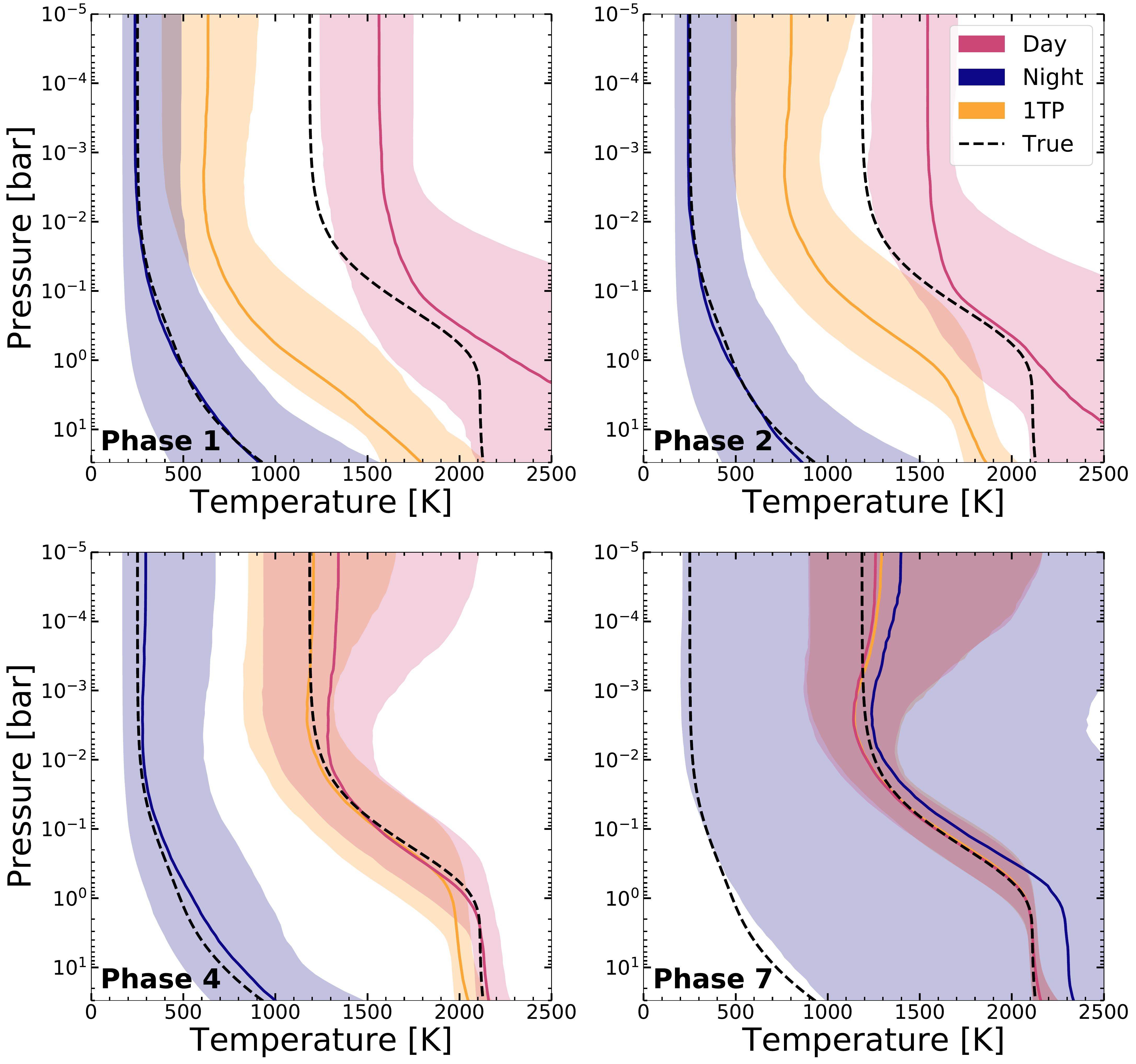}
  \caption{Temperature-pressure (TP) profiles for simulated HST+Spitzer data. We selected phases 1, 2, 4, and 7 to show the change in TP profile constraint as a function of phase. In each panel, the dashed line represent the true input profiles for the day and night sides. The retrieved 2TP-Crescent profiles are in blue (night) and red (day). The retrieved 1TP profiles are in yellow. For each distribution, we show the median profile in a solid line, surrounded by the $2\sigma$ spread in profiles. 1TP-retrieved profiles fall in between the true day and true night profiles, shifting toward hotter temperatures until reproducing the true day profile at secondary eclipse. The 2TP-Crescent model provides constraints on the night profiles for most of the orbit, until secondary eclipse where there is negligible night side emission. There is a preference for hotter temperatures for the day side at phases closer to transit (more of the night side visible), but once we reach quarter phase and above, the day side profile is accurately constrained.}
  \label{fig:hst_tp}
\end{figure}

Figure \ref{fig:hst_tp} shows the progression of the retrieved pressure-temperature profiles from the two models as a function of phase. For 1TP, as the phase gets closer to secondary eclipse, the retrieved profile matches more closely with the input dayside profile. The 1TP retrieved profiles are biased towards hotter temperatures at phases near primary transit ($<$ phase 3). This is because the 1TP profile is attempting to strike a balance between the nightside and dayside fluxes, as discussed in \citet{Feng2016}. When implementing the 2TP-Crescent model, the retrieved day and night profiles more-or-less retrieve the input profiles, with mild bias for phases before phase 3, where more of the nightside TP profile is present. At secondary eclipse, there is no emission signal from the night, so the TP profile for the night at this phase is completely unconstrained, effectively filling out the prior.

\subsection{Application to the Observed WASP-43b Data Set}
\label{subsec:w43-results}
Here, we consider the retrieval outcome using actual observations of WASP-43b from \citet{Stevenson2014,Kreidberg2014,Stevenson2017} using HST+Spitzer. We consider results from the 1TP, 2TP-Crescent, and 2TP-Free models. The two different 2TP retrieval implementations allow for comparison to simulated data results and exploration of asymmetry in the atmosphere as a function of phase. The 2TP-Free model enables the latter because the dayside contribution, as parameterized by $f_{\rm day}$, is a retrieved quantity. 

As Figure \ref{fig:sigmaCompare} (middle panel, outer ring) shows, {phases 1-3 and 10-13} are moderately in favor of the 2TP-Crescent model {when compared to the 1TP model. Phase 4 is weakly in favor.} The inner ring in the same panel compares the 1TP model and the 2TP-Free model. For this case, phases 1-3, 10, and 13 are moderately in favor of the Free model; phases 11 and 12 are strongly in favor; {phase 4 is weakly in favor}. 

\begin{figure*}[!tbp]
  \centering
  \includegraphics[width=0.9\textwidth]{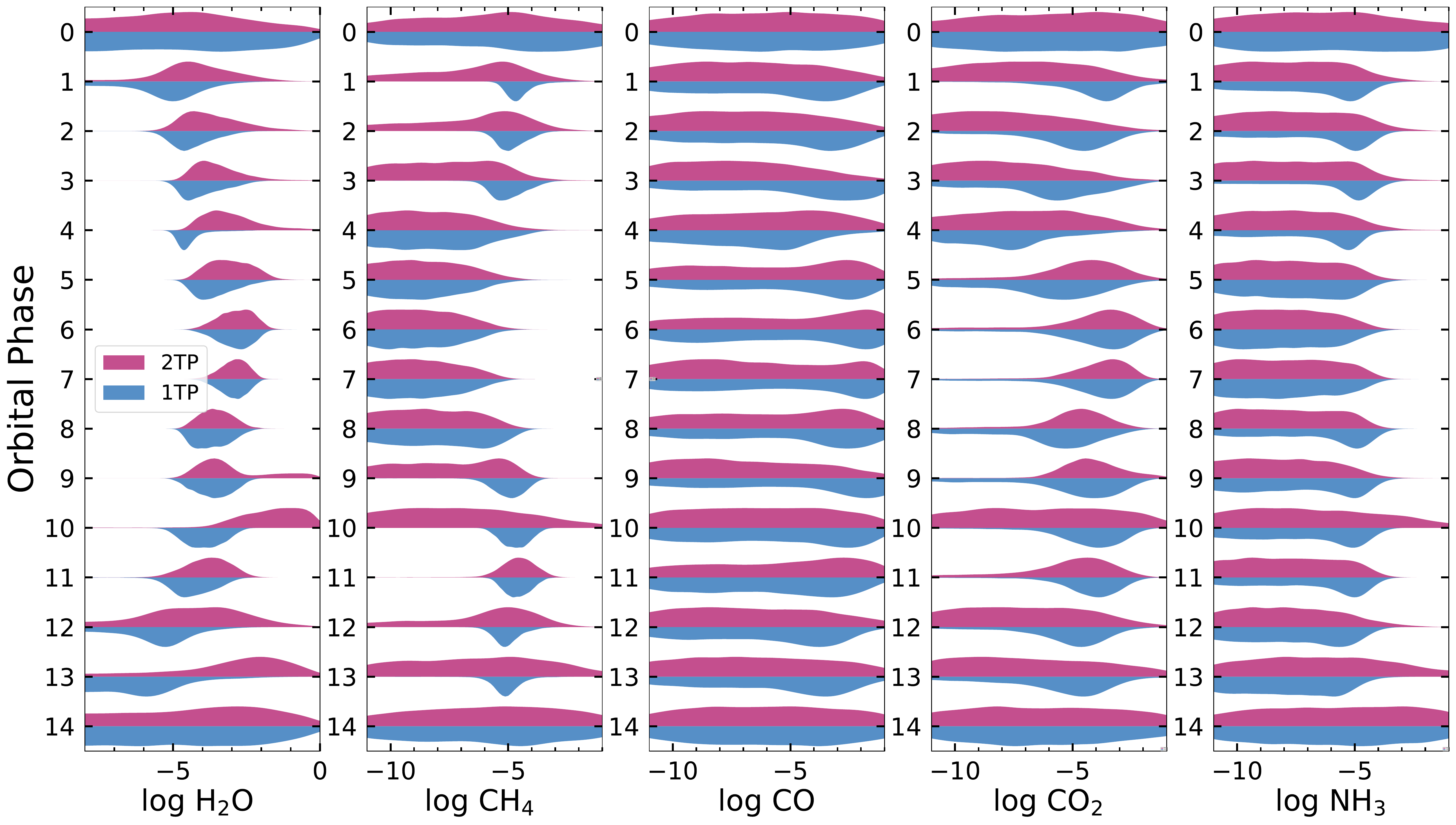}
  \caption{Abundance vs. phase results from WASP-43b data for \water, \methane, CO, \cotwo, and \ammonia\ for the 1TP model (blue) and the 2TP-Crescent model (dark pink). For each panel, we plot {the kernel density estimation of} the posterior probability distribution of the log of the molecule's mixing ratio as a function of orbital phase. We see artificially tight constraints of \methane\ at several phases when using the 1TP model. With 2TP-Crescent, \methane\ at phase 11 is also constrained. However, considering the constraints (of lack thereof) of all the phases can help identify potential outlier distributions. \water\ constraints from the two models are consistent, with similar increases in estimates from transit to secondary eclipse {(phases 0 to 7)}. {Beyond secondary eclipse, \water\ appears to be discrepant at phases 10 and 13.} There is no constraining power within the data sets for CO, {while upper limits can be placed on \ammonia.} \cotwo\ is largely unconstrained with the exception of phases near secondary eclipse. In some cases the 1TP model results in overconstrained abundances relative to the 2TP-Crescent model. \cotwo\ constraints are challenging to interpret due to the 1-to-1 degeneracy with CO as a result of the overlapping bands over the 4.5$\mu$m Spitzer point.}
  \label{fig:w43_abunds}
\end{figure*}

Figure \ref{fig:w43_abunds}, like Figure \ref{fig:hst_abunds}, summarizes the molecular abundance constraints for the 1TP and 2TP-Crescent scenarios, in this case for all 15 phases in the \citet{Stevenson2017} WASP-43b data set.   

We find that the retrieved distributions for all gases mostly resemble the trends seen from the simulated data set (Figure \ref{fig:hst_abunds}). There are no substantial biases in the constraints on \water, CO, \cotwo, or \ammonia, though the water abundance appears to increase at phases between first and third quarter (e.g., around secondary eclipse)

\water\ is the only well-constrained (i.e., bounded) species. Overall, there is no significant difference in the posteriors between the 1TP and the 2TP-Crescent models for \water, \ammonia, CO, or \cotwo.  Although, at phase 10 (almost at third quarter), the 2TP \water\ posterior indicates slightly elevated values than the rest of the orbit.

We see the same constrained distributions for \methane\ that are only present under the 1TP model as seen in the simulated data set. Once again, the affected phases are when the visible hemisphere is dominated by the night side. The only phase that does not have an upper limit distribution under the 2TP model is phase 11.  In this case, we see a well-constrained posterior consistent with what the 1TP finds. {We also find that most of the phases where there is evidence for the 2TP-Crescent model in Figure \ref{fig:sigmaCompare} have biased \methane\ posteriors under the 1TP model. Of these, phase 4, with weak evidence, places an upper limit on \methane.}

\begin{figure}[!tbp]
  \centering
  \includegraphics[width=0.45\textwidth]{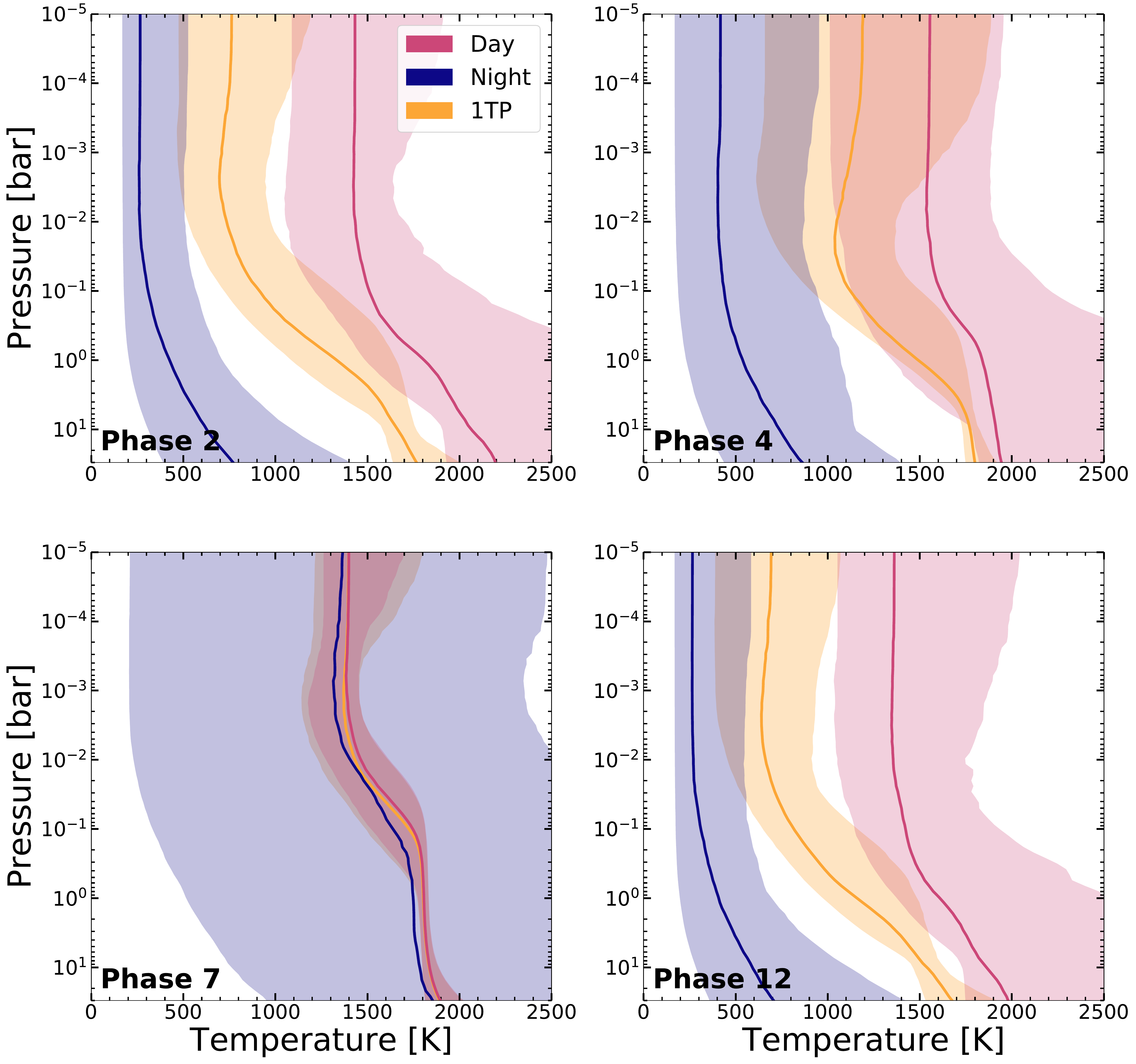}
  \caption{Temperature-pressure (TP) profiles for HST+Spitzer data of WASP-43b. Phases 2, 4, 7, and 12 are shown to illustrate the constraint behavior with phase. The retrieved 2TP-Crescent profiles are in blue (night) and red (day). The retrieved 1TP profiles are in yellow. For each distribution, we show the median profile in a solid line, surrounded by the $2\sigma$ spread in profiles based on reconstructed from random posterior parameter draws. Phases 2 and 12 are symmetric in the orbit (just after and just before transit, respectively), resulting in similar retrieved profiles under both models. The retrieved 1TP profile overlaps perfectly with the 2TP-Crescent dayside profile at secondary eclipse (phase 7) as there is no contributing flux from the nightside. Retrieved day and night profiles from the 2TP-Crescent model are relatively similar from phase to phase, further evidence of a large day-night temperature contrast in the atmosphere.}
  \label{fig:w43_tp}
\end{figure}

Figure \ref{fig:w43_tp} summarizes the retrieved TP profiles for select phases, as in Figure \ref{fig:hst_tp}. The behavior of the retrieved profiles over the orbit resemble what is seen from simulated data. We note that the overlap between the 1TP model profiles and the day side profiles from the 2TP-Crescent model at phase 4 is smaller than in the simulated case. The day side profiles also appear more isothermal, with a smaller temperature gradient through photospheric pressures.  

\begin{figure*}[!tbp]
  \centering
  \includegraphics[width=0.9\textwidth]{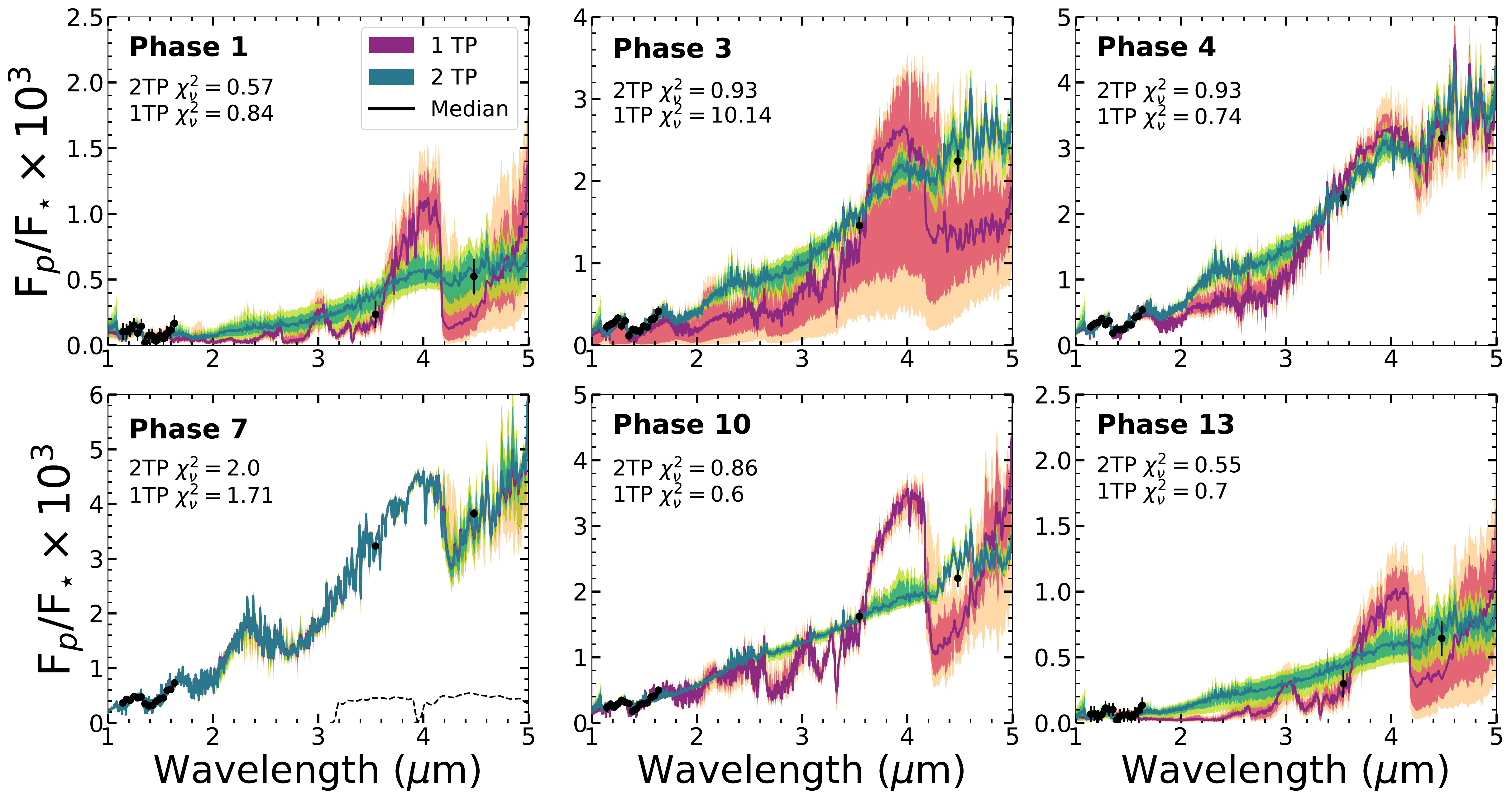}
  \caption{WASP-43b data (HST+Spizter) and high-resolution spectra generated with random posterior draws from the retrieval. Shown here are the spectra for phases 1, 3, 4, 7, 10, and 13. We include corresponding $\chi^2_\nu$ values for the 1TP and 2TP(-Crescent) models. In the panel of phase 7, we overplot the Spitzer 3.6$\mu$m and 4.5$\mu$m filter transmission. 1TP spectra are in magenta while 2TP-Crescent spectra are in green. For each set of model spectra, we plot the median, $1\sigma$, and $2\sigma$ contour. The 1TP model struggles to fit the 4.5$\mu$m Spitzer point at more crescent phases (dominated by night side). The 2TP-Crescent model spectra look more featureless in comparison at these phases, reflecting the corresponding unconstrained posteriors of the atmospheric gases. At phases closer to secondary eclipse, data between 2 and 3$\mu$m are needed to separate the two models. {At secondary eclipse, the two models agree as expected.}}
  \label{fig:w43_spectra}
\end{figure*}

In Figure \ref{fig:w43_spectra}, we show representative model fits for both the 1TP and 2TP-Crescent scenarios. The fits of these two models are statistically similar {(e.g., at phase 7, 1TP model's $\chi^2_\nu = 1.71$, while 2TP-Crescent's $\chi^2_\nu$ = 2.00)} for phases around secondary eclipse (6 - 8). There is little ``nightside'' contribution in these cases, permitting an adequate representation with a single TP. At most phases shown, the 1TP model requires significantly larger abundances (sometimes for \ammonia\ or \methane\ depending on the phase) and a steeper temperature gradient than the 2TP-Crescent model, resulting in much more spectral contrast, which yields deep absorption features, to fit the spectra and photometry. The 2TP fitted spectra, in comparison, have less spectral contrast overall. Interestingly, while Phases 4 and 10 are in theory geometrically symmetric, the overall flux at phase 10 is lower and the difference between the two models is more noticeable at that phase. This is possibly due to the presence of an offset hotspot, leading to asymmetry between phases before and after secondary eclipse and a difference in fitted model parameters. 

\subsubsection{2TP-Crescent vs. 2TP-Free}
\label{2tpfree}
As introduced in Section \ref{subsec:investigate}, we consider several ways to retrieve an atmosphere with two contrasting thermal profiles. For observed data in particular, we are interested in identifying longitudinal asymmetry in the atmosphere (e.g., due to a hot spot offset). Using the 2TP-Free model can inform us about potential inhomogeneity, and we may study how the interpretation of an atmosphere changes when there is more flexibility in geometry. Here we highlight the differences between the 2TP-Crescent model and the 2TP-Free model on the WASP-43b data set. 

As described in Section \ref{subsec:investigate}, the 2TP-Crescent model uses the geometry described in Figure \ref{fig:schematic}. The 2TP-Free model uses a linear combination of fluxes from the day side and fluxes from the night side, parameterized with $f_{\rm day}$, which accounts for the fraction contributed by the day side. There is no assumption of geometry or symmetry in the free model. The $f_{\rm day}$ parameter determines how much of the final spectrum is contributed by the hot TP profile, while the remaining flux is then attributed to the cool TP profile. We have the ability to see whether the retrieved $f_{\rm day}$ from a certain phase's data set is different from the corresponding value based on Equation \ref{eq:illum}

\begin{figure}[!tbp]
  \centering
  \includegraphics[width=0.45\textwidth]{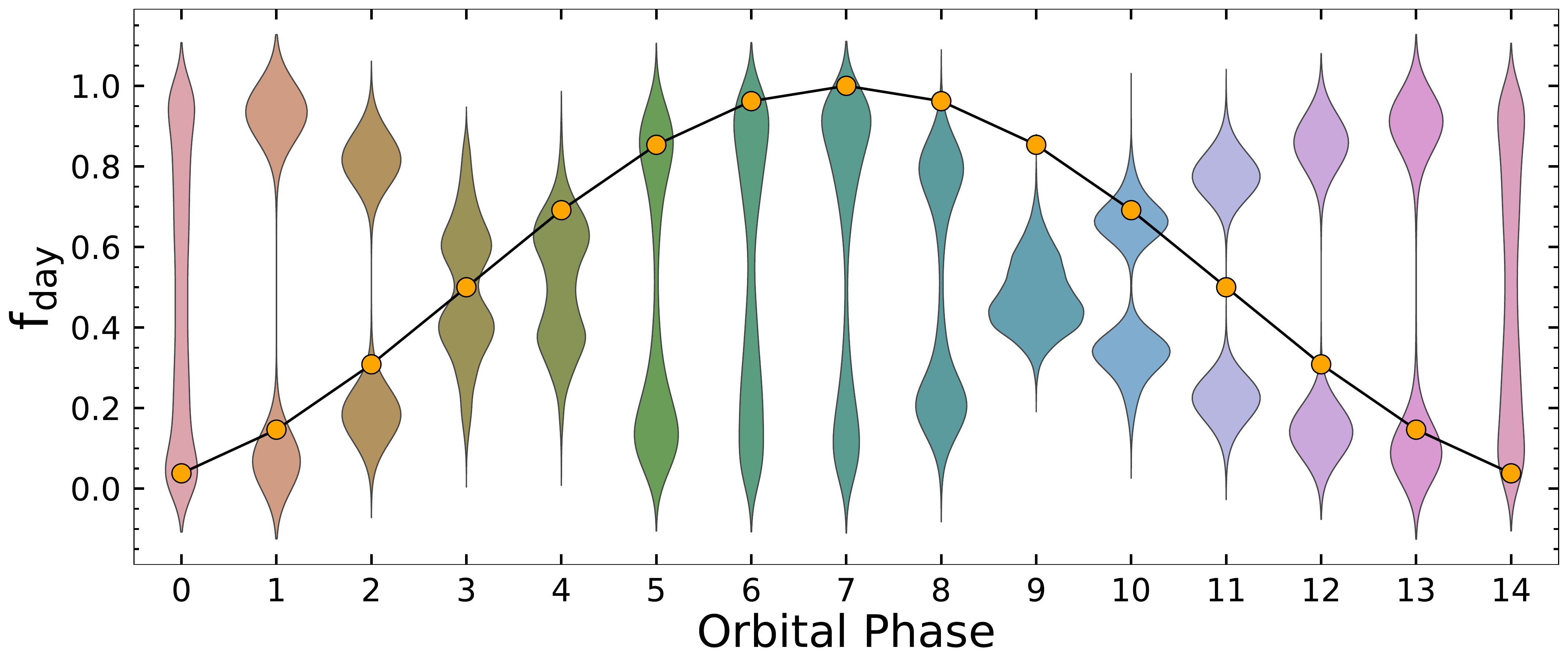}
  \caption{Posterior distribution of $f_{\rm day}$ from the 2TP-Free model using WASP-43b data. The distributions are bimodal due to the fact that we imposed no geometric information ({day and night fraction are interchangeable for symmetric phases}). Overplotted (orange circles connected with black line) is the expected emitting fraction for each phased based on Equation \ref{eq:illum}. These expected values correspond to the total contribution from the day side in the 2TP-Crescent model. Phases 2, 9, 11, and 12 have posteriors constraining lower values than the expected, suggesting a preference for lower temperatures and less contribution from the day profile. }
  \label{fig:w43_fday}
\end{figure}

In Figure \ref{fig:w43_fday}, we show the posterior distributions as a function of phase for the parameter $f_{\rm day}$ from the 2TP-Free model. In particular, we note the lower-than-expected value at phase 2, 9, 11, and 12. This suggests a preference for lower flux from the day side or hotter temperatures and more flux from the night side or lower temperatures. Furthermore, this preference is seen mostly past secondary eclipse. WASP-43b has a known hot spot offset such that the maximum flux occurs before secondary eclipse \citep{Stevenson2017}; the behavior in $f_{\rm day}$ as a function of phase is thus evidence for the offset. We can also examine phases 5 through 7 (right before to during secondary eclipse) and see that the $f_{\rm day}$ posteriors appear similar rather than finding higher values at secondary eclipse. 

\begin{figure}[!tbp]
  \centering
  \includegraphics[width=0.45\textwidth]{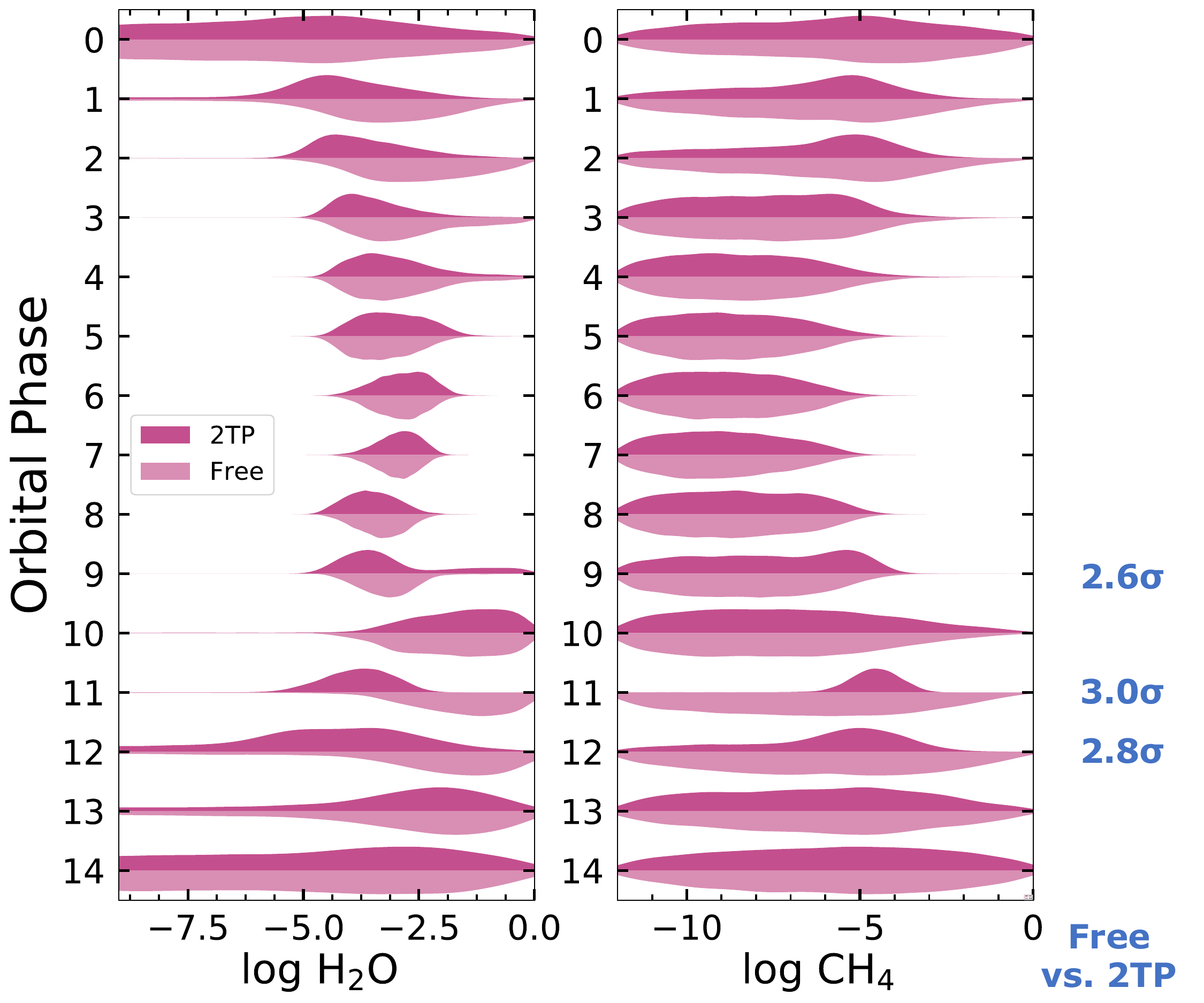}
  \caption{Abundance vs. phase results from WASP-43b data for \water\ (left) and \methane\ (right) for the 2TP-Crescent model (dark pink) and the 2TP-Free model (light pink). For each panel,we plot {the kernel density estimation of} the posterior probability distribution of the log of the molecule's mixing ratio as a function of orbital phase. A noticeable difference is how 2TP-Crescent's constraint of \methane\ at phase 11 becomes a non-detection with 2TP-Free. The \water\ distributions from phases 10 - 13 with the 2TP-Free model look more similar to one another but these values are larger than estimates from the rest of the orbit. {On the far right of the figure, we annotate the detection significance of the 2TP-Free model over the 2TP-Crescent model at phases where the significance $\sigma_{\rm Free} > 2$. Here, we see phases 9, 11, and 12 showing weak to moderate evidence in favor of the 2TP-Free model; see Figure \ref{fig:sigmaCompare} for definitions.}}
  \label{fig:w43_2tpfree_abunds}
\end{figure}

Next, we consider how the additional geometric flexibility impacts atmospheric inference. Figure \ref{fig:w43_2tpfree_abunds} shows the posterior distributions for \water\ and \methane\ for these two models. CO, \cotwo, and \ammonia\ are virtually identical and these comparisons have been left off the figure. The most significant differences for \water\ and \methane\ occur on the night side as seen after secondary eclipse, in particular at phases 11 and 12. These correspond to some of the phases that returned lower-than-expected day side flux contribution in the 2TP-Free model. The 2TP-Crescent model constrains \methane\ at these two phases. The 2TP-Free model, on the other hand, shows no detection of the molecule.  

In terms of \water, the 2TP-Free model shows similar posterior distributions from phases 10 through 13. These values are consistent with the findings at phases 10 and 13 under the 2TP-Crescent model. However, the 2TP-Free values are elevated compared to the posteriors for phases 11 and 12 with 2TP-Crescent. These are the two phases that became non-detection under 2TP-Free for \methane. {Upon examining the retrieved TP profile structure, we find that the day and night profiles from the 2TP-Crescent and 2TP-Free models are similar at phases 10 and 13. In addition, the 2TP-Free model retrieves the expected $f_{\rm day}$ for phases 10 and 13, unlike e.g., phase 11, such that the overall day-night flux contribution is similar to what is used under 2TP-Crescent. Thus, the two models return similar retrieved abundances at phases 10 and 13 as well in order to fit the data.}

{We also examine the Bayesian evidence of the 2TP-Free model over 2TP-Crescent model;} the 2TP-Free model has one additional free parameter $f_{\rm day}$ over the 2TP-Crescent model. {Phase 9 has a detection significance of $2.6\sigma$, weakly in favor of 2TP-Free. Phases 11 and 12 each has a detection significance of $3\sigma$ and $2.8\sigma$ respectively, which is moderately in favor of 2TP-Free (see Figure \ref{fig:w43_2tpfree_abunds}). All other phases have a detection significance $<2\sigma$, or inconclusive evidence for 2TP-Free.} {Furthermore, when we separately looked at the 2TP-Free model performance over the 2TP-Crescent model for simulated HST+Spitzer data, we found that all phases showed inconclusive evidence for 2TP-Free.} As a result, the more justified 2TP model appears to be the 2TP-Crescent model {for HST+Spitzer observations}. 

\begin{figure*}[!tbp]
  \centering
  \includegraphics[width=0.95\textwidth]{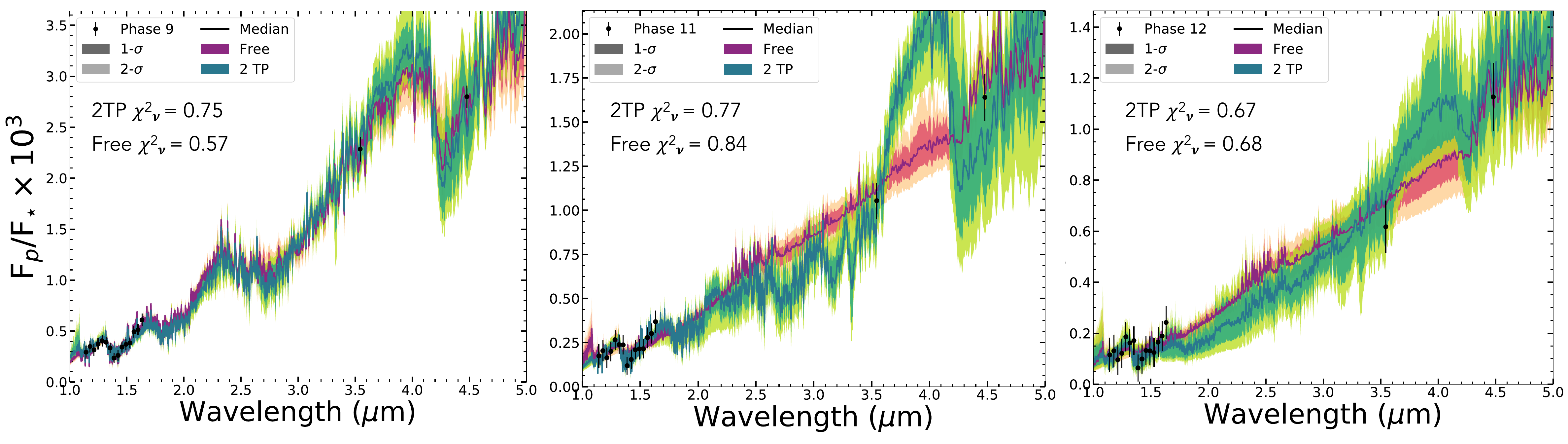}
  \caption{{WASP-43b data (HST+Spizter) and high-resolution spectra generated with random posterior draws from 2TP-Crescent (2TP, green) and 2TP-Free (Free, magenta) models. Shown here are the spectra for phases 9, 11, and 12, which are the phases with weak to moderate evidence for the 2TP-Free model over the 2TP-Crescent model. We include corresponding $\chi^2_\nu$ values for the 2TP-Crescent and 2TP-Free cases. In the panel of phase 7, we overplot the Spitzer 3.6$\mu$m and 4.5$\mu$m filter transmission. For each set of model spectra, we plot the median, $1\sigma$, and $2\sigma$ contour. There is noticeable difference in the fits for the Spitzer photometric points for phases 11 and 12 in particular.}}
  \label{fig:free2tp_spec}
\end{figure*}

The 2TP-Free model, however, is able to account for existing asymmetries in the phase curve. In this regard, we note that the 2TP-Crescent model can be modified in the future to retrieve for an arbitrary phase angle such that the dayside contribution is not pre-determined based on geometry. {We visualize the spectra generated with the parameters retrieved at phases 9, 11, and 12 in Figure \ref{fig:free2tp_spec}. Phases 11 and 12 have discrepant fits to the Spitzer photometric points, and differences between $2-3\mu$m. These fits manifest the dissimilar posterior abundances for e.g., \water\ and \methane\ seen in Figure \ref{fig:w43_2tpfree_abunds} at these phases, as well as the inclusion of $f_{\rm day}$ under 2TP-Free which adds a degree of freedom in capturing the overall thermal structure and how much hot or cool flux contributes to the final spectrum.}

\subsection{Simulated JWST data}
\label{subsec:jwst-results}
From the above analysis on both simulated and true HST+Spitzer data, we found that substantial abundance and temperature biases exist when applying a single TP profile to several planetary phases. {An important trend emerged: standard Bayesian nested modeling tools are able to rule out the simpler model which resulted in biased results.} We now present the effects of the 1TP vs 2TP-Crescent assumption on simulated JWST data with the anticipation that {this trend will become more obvious}. 

A look back at the rightmost panel of Figure \ref{fig:sigmaCompare} shows that all phases (with the exception of secondary eclipse due to the complete lack of nightside contribution) demonstrate a clear preference for the more complex 2TP-Crescent model. We describe this result here. 

\begin{figure*}[!tbp]
  \centering
  \includegraphics[width=0.9\textwidth]{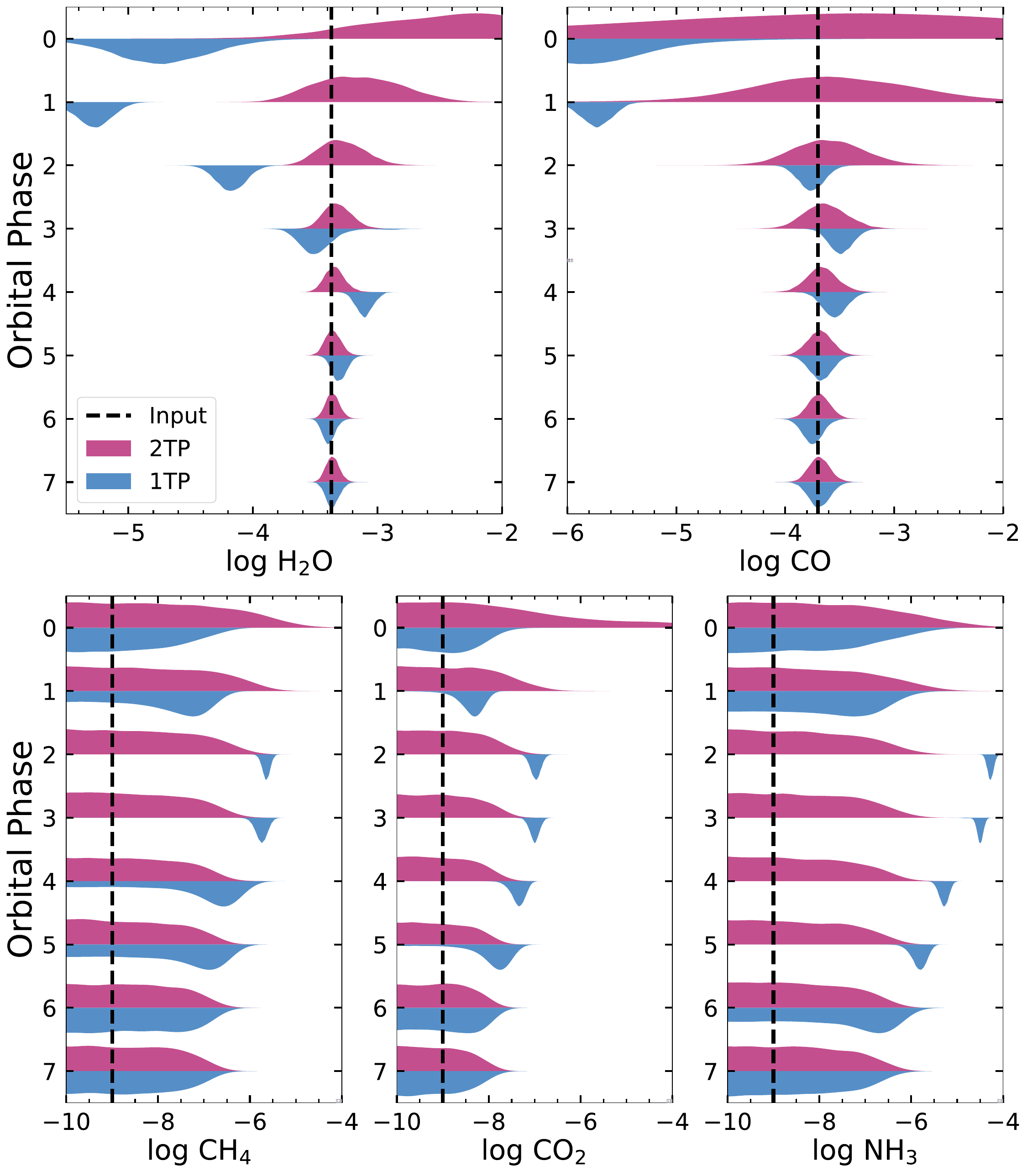}
  \caption{Abundance vs. phase results from \textit{JWST} simulated data for \water, \methane, CO, \cotwo, \ammonia\ for the 1TP model (blue) and the 2TP-Crescent model (dark pink). For each panel, we plot {the kernel density estimation of} the posterior probability distribution of the log of the molecule's mixing ratio as a function of orbital phase. For simulated data, we only consider half an orbit (transit to secondary eclipse), or eight orbital steps. For each molecule, we indicate its input abundance value with the vertical dashed line. The 1TP model produces constrained but bias posteriors for all molecules at multiple phases. Most of them have incorrect estimates for half the orbit. With the 2TP-Crescent model, we can get well-constrained and accurate estimates of \water\ and CO. We have upper limits for the remaining molecules, which do not have large input values to begin with. }
  \label{fig:jw_abunds}
\end{figure*}

The broad (1-10$\mu$m) and higher resolution ($R=100$) simulated JWST data, {combining a single transit each in NIRISS, NIRCam, and MIRI LRS,} provides ultra-precise constraints on the gas mixing ratios when using the correct model. This high quality data permits us to readily disprove the 1TP hypothesis at all phases ($>5\sigma$), at least in this scenario of a phase curve from a planet with large day-night temperature contrast. Figure \ref{fig:jw_abunds} shows the retrieved molecular abundance constraints as a function of phase. Phase 7, or secondary eclipse, is the only phase where the 1TP model does not produce a bias, as expected. Phase 0 is not informative for abundances for either model, owing to the overall low night-side flux. For many phases, strong abundance biases across all the gases persist under the incorrect 1TP profile model. 

While for the simulated HST data there is negligible difference in \water\ inference between the two models, all of the 1TP posteriors here for \water\ miss the input value until phase 6. The problem is worsened by the precision in the biased posterior distributions. For example, the 1TP \water\ constraint in phase 1 is $\log \rm H_2O = -5.28^{+0.14}_{-0.15}$ when in fact the true mixing ratio is $\log \rm H_2O = -3.37$ (or a 1TP bias of $12.7\sigma$). The 2TP-Crescent (true model) at phase 1, however, results in a less precise ($^+_-0.35$), but more accurate (unbiased) constraint. Unsurprisingly, the water precision improves ({up to $^+_-0.05$ dex at secondary eclipse}) as more dayside is visible due to the higher emission feature signal to noise. We see up to a factor of 5 improvement in precision over the phase 7 posterior with 2TP-Crescent model using simulated HST data. 

The 1TP CO constraints present less of a bias than \water, as seen in Figure \ref{fig:jw_abunds}.  CO begins to largely deviate from the truth between phases 0 and 4. Although, phase 2 provides an unbiased yet over-constrained abundance compared to the 2TP-Crescent constraint. By phase 7, we can constrain CO using the 2TP-Crescent model to a $1\sigma$ of $\sim 90$ppm ($\log \rm CO = -3.69^{+0.1}_{-0.09}$). 

The poor performance of the 1TP model is further evident in \methane, \cotwo, \ammonia\ estimates. We find biased posteriors that are well constrained to 1/10th of an order of magnitude but their median values are several orders of magnitude away from the truth. \methane, for instance, suffers from bias at phases 2-4. \cotwo\ is biased at phases 1 through 5, and \ammonia\ from phase 2 to 5. Meanwhile, the 2TP-Crescent model only detects upper limits for these molecules.    

\begin{figure}[!tbp]
  \centering
  \includegraphics[width=0.45\textwidth]{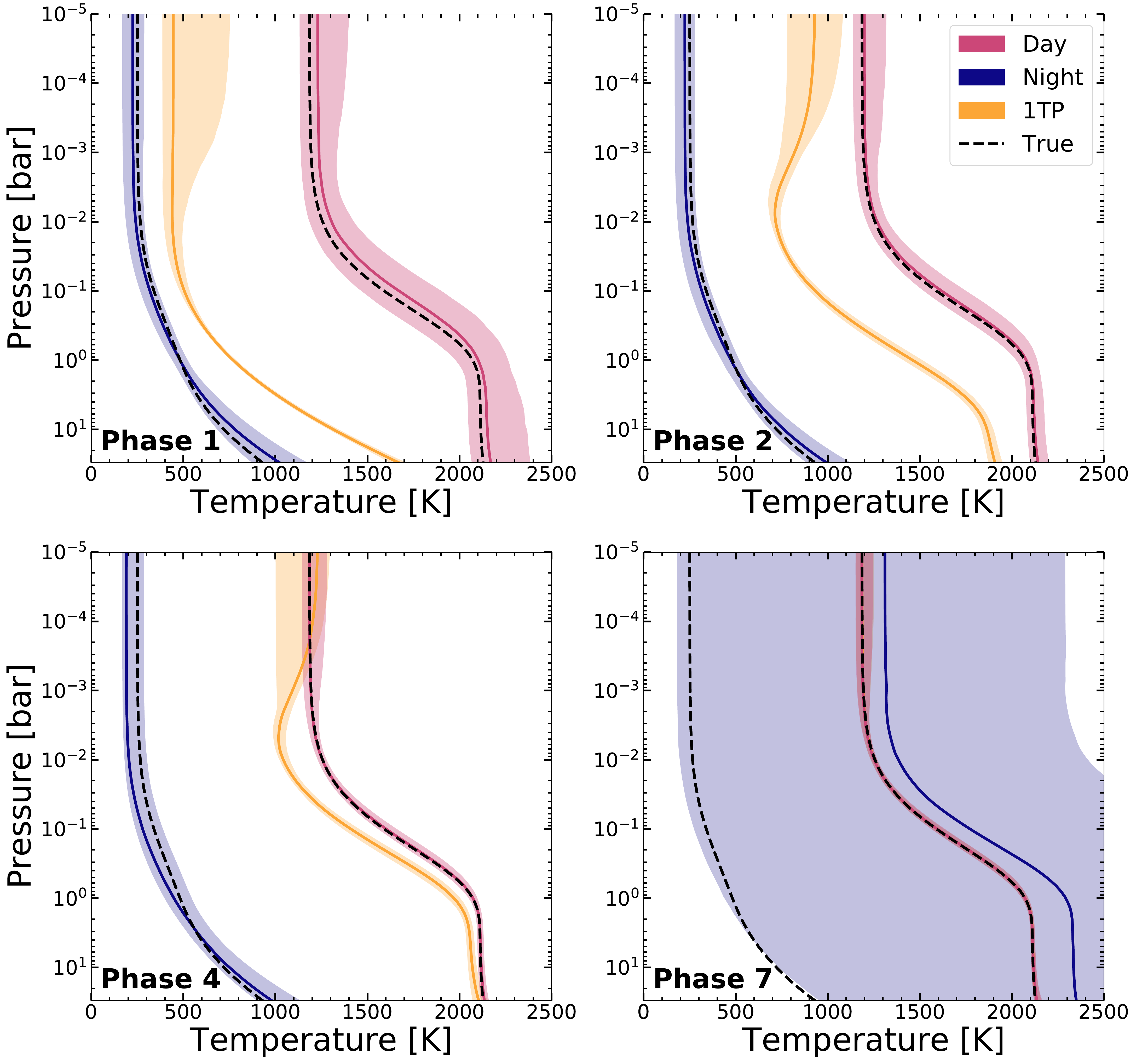}
  \caption{Temperature-pressure (TP) profiles for simulated JWST data. We selected phases 1, 2, 4, and 7 to show the change in TP profile constraint as a function of phase. In each panel, the dashed line represent the true input profiles for the day and night sides. The retrieved 2TP-Crescent profiles are in blue (night) and red (day). The retrieved 1TP profiles are in yellow. For each distribution, we show the median profile in a solid line, surrounded by the $2\sigma$ spread in profiles based on reconstructed random posterior draws. For certain phases, the 1TP profiles appear to have a temperature inversion. The 1TP profiles are close to the day-side profile as early as phase 4 (half day, half night). The 2TP profiles for day and night are accurate and precise. }
  \label{fig:jw_tp}
\end{figure}

Figure \ref{fig:jw_tp} summarizes the retrieved TP profiles based on the two models. While overall trends are similar to what we see in simulated HST data, we find much more precisely constrained profiles across the orbit, by a factor of several better than with current data. Consequently, we note the presence of artificial temperature inversions at phases 2 and 4 under the 1TP model between $10^{-1}$ bar and $10^{-4}$ bar. At other phases, such as phase 3 and 5 (not shown in Figure \ref{fig:jw_tp}), we do not find this phenomenon. This is yet another example of false conclusions that could arise from the overly simplistic 1TP profile assumption.


\begin{figure}[!tbp]
  \centering
  \includegraphics[width=0.45\textwidth]{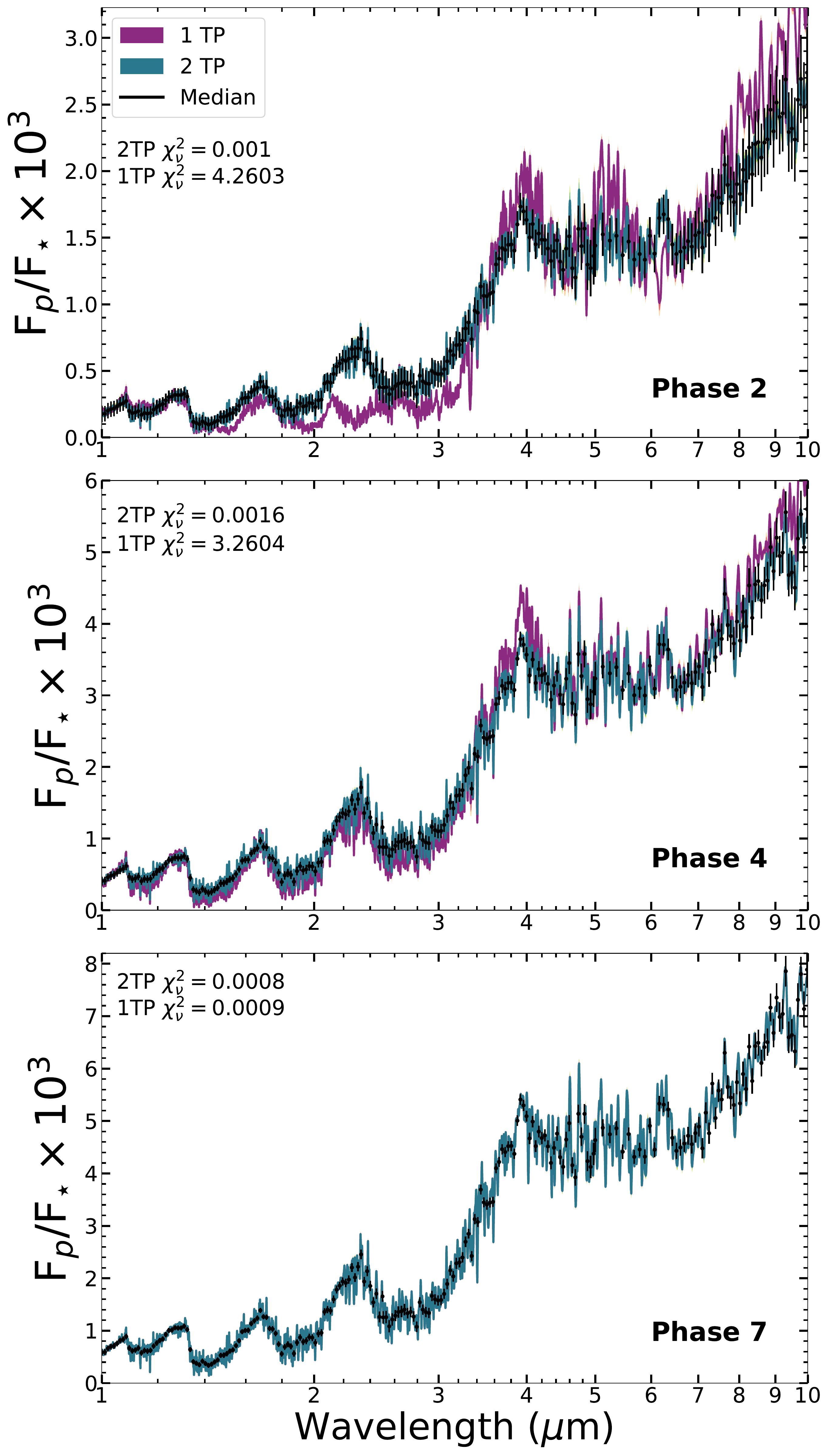}
  \caption{Simulated \textit{JWST} data and high-resolution spectra generated with random posterior draws from the retrieval. Shown here are the spectra for phases 2, 4, and 7. 1TP spectra are in magenta while 2TP-Crescent spectra are in green. For each set of model spectra, we plot the median, $1\sigma$, and $2\sigma$ contour. We include corresponding $\chi^2_\nu$ values for the 1TP and 2TP(-Crescent) models, which can be small because random noise is not included. The \textit{JWST} results are so precise that the contours are difficult to see. The 1TP spectra do not fit the majority of the data points at phases besides secondary eclipse. }
  \label{fig:jw_spectra}
\end{figure}

Figure \ref{fig:jw_spectra} shows the simulated data and fitted spectra from the two models. As anticipated based on the abundance inference, the 1TP model produces poor fits. At phase 2, we see that the model spectra miss the data points between $1.4-3.2\mu$m, with a reduction in flux with respect to the data. Then, between $3.6-10\mu$m, we see elevated flux. The quality of fit from the two models is easily distinguishable due to the precise $1\sigma$ and $2\sigma$ contours, and the 1TP model is not able to properly fit the data until just before secondary eclipse.

\subsection{Joint Phase Retrieval}
\label{subsec:simretrieve}
The previous phase-by-phase analyses are useful for determine phase-dependent properties like abundance variations.  They also demonstrated the possibility of strong biases when using too simplistic of a model. Here we explore the feasibility of a joint phase curve retrieval whereby we simultaneously retrieve upon all phases, locking certain properties at each phase. The goal is to determine if improved precision can be obtained on atmospheric properties that are expected to be uniform with phase. { This assumption is motivated by coupled GCM-chemical kinetics investigations that suggest species like water and CO are expected to be homogenized both horizontally and vertically due to the relatively short transport timescales relative to the kinetic timescales \citep{coopershowman2006, agundez2012}} This approach in a sense would be an intermediate step before employing full ``3D GCM-retrievals''.  

We explore the joint phase curve retrieval on three cases: both the simulated and \citet{Stevenson2017} HST+Spitzer WASP-43b phase curve data as well as simulated JWST phase curve data. In all three observational scenarios, we explore two separate 2TP modeling assumptions:  the first is our new geometric method where limb darkening is appropriately accounted for per Gaussian quadrature annulus (2TP-Crescent), and the second is 2TP-Fixed, where we apply parameter $f_{\rm day}$, a set value based on Equation \ref{eq:illum}, to account for phase-dependent fluxes. 

\begin{figure*}[!tbp]
  \centering
  \subfigure[Simulated HST+Spitzer: Constraints from joint retrievals]{\includegraphics[width=0.31\textwidth]{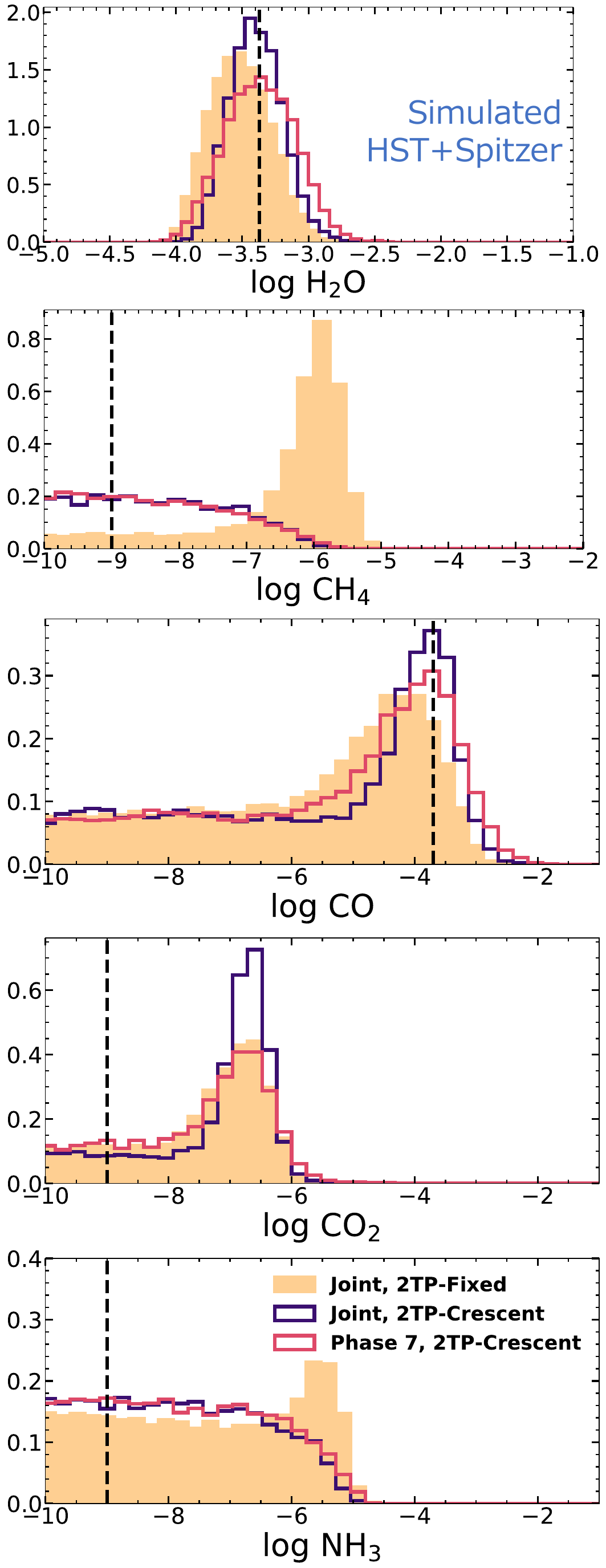}}\hfill
  \subfigure[WASP-43b: Constraints from joint retrievals]{\includegraphics[width=0.31\textwidth]{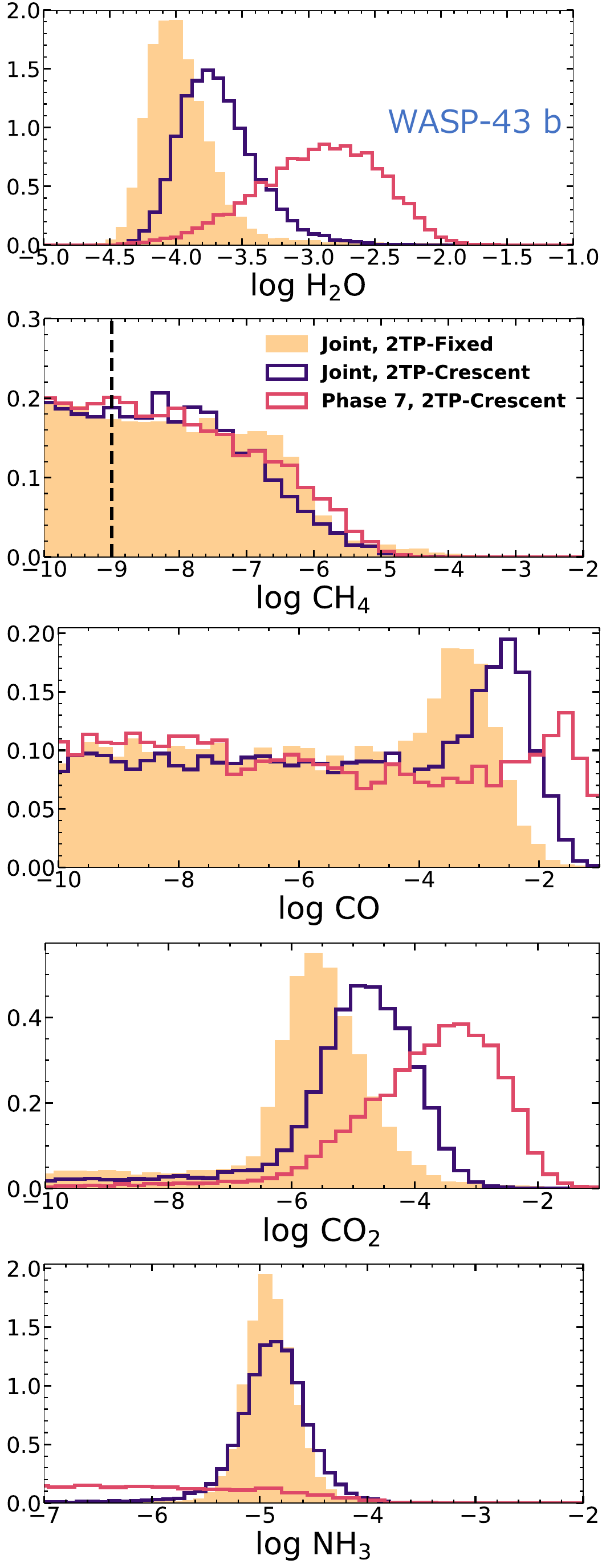}}\hfill
  \subfigure[JWST: Constraints from joint retrievals \label{fig:all_jw}]{\includegraphics[width=0.31\textwidth]{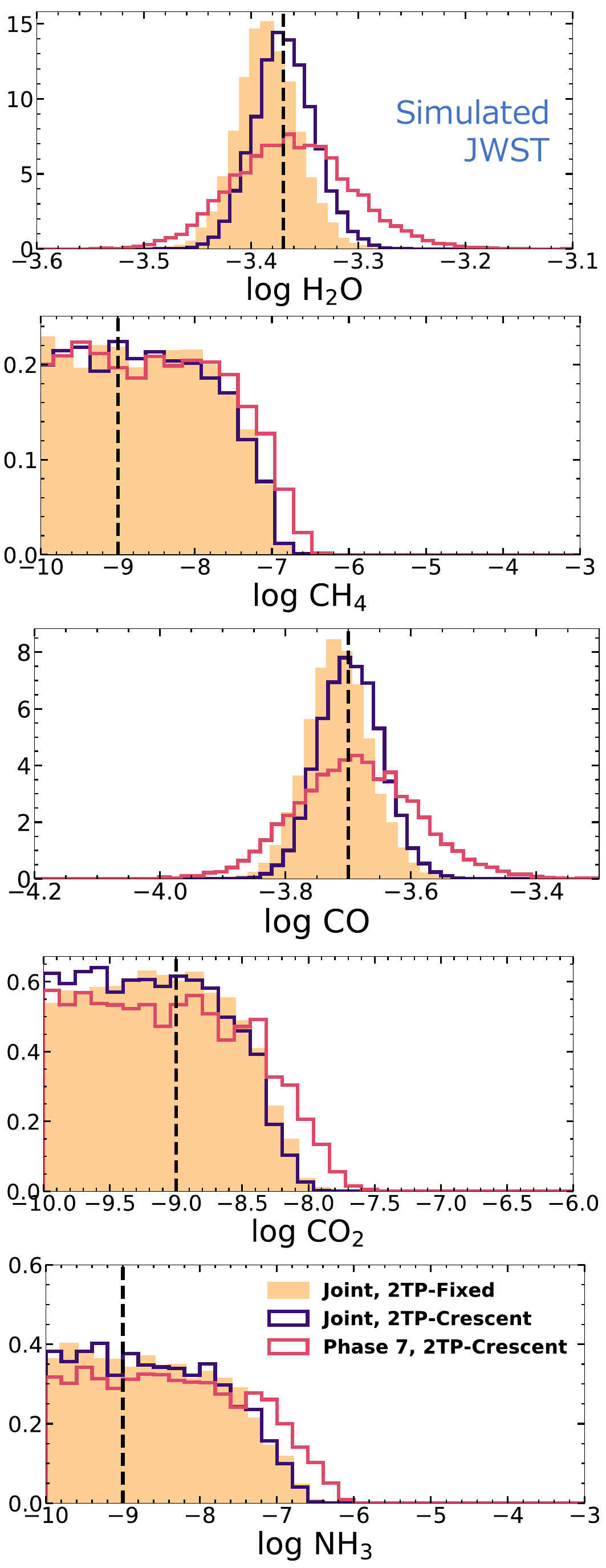}}
  \caption{{Normalized probability density vs. abundance from joint retrievals.} Left: Constraint distributions of the five retrieved molecules from simulated HST+Spitzer data with: (1) averaged posterior from the phase-by-phase retrievals from Section \ref{subsec:hst-results}, (2) joint retrieval of all phases using the 2TP-fixed model,(3) joint retrieval of all phases using the 2TP-Crescent model. Dashed line indicates the input value for each molecule. Middle: Posterior distributions of the same cases using WASP-43b data. The joint retrieval is able to return more precise distributions; in some cases, however, the advantage of combining multiple data sets also enhances bias in the result. Right: Posterior distributions of the same cases using simulated JWST data. Distributions from jointly-done retrievals indicate stronger, more precise detection. 2TP-Fixed and 2TP-Crescent approaches yield similar results.}\hfill
  \label{fig:all_hst}
\end{figure*}

In these examples, we assume that the gas mixing ratios are the phase-independent quantities with only the contribution of the temperature profiles changing with phase. Figure \ref{fig:all_hst} summarizes the abundance constraints under these conditions. We also include the constraints from Sections \ref{subsec:w43-results} and \ref{subsec:jwst-results} at phase 7 (secondary eclipse) for comparison. 

For most molecules, the joint retrieval shows similar results to the phase 7 retrieval in the case of simulated data. This is expected as the abundances are designed to remain constant with phase. The only case where the two joint methods differ significantly is for \methane. While the joint 2TP-Crescent retrieval does not retrieve phantom elevated amounts of \methane, the joint 2TP-Fixed retrieval suffers from this bias, {although we do note the extended tail to the distribution}. CO is {weakly} detected at the input value while having a large tail to the distributions that does not rule out lower values. Both models also {weakly detect} \cotwo\ for the simulated data {with an unbounded distribution tail}; however, we attribute these to the correlation seen between CO and \cotwo\ based on Spitzer data points. 

For WASP-43b phase curve data, we see larger discrepancies between the two joint approaches as well as the results from the individual phase 7 retrieval. In contrast to the simulated case, it appears that the secondary eclipse conditions are not representative of the planet over the orbit. We do not see any \methane\ bias for either model. The CO distributions peak at different values, unlike in the simulated data results. \cotwo, which is constrained again, exhibits similar results, with secondary eclipse peaking at the higest value, followed by joint 2TP-Crescent and then joint 2TP-Fixed; this is demonstration of the CO-\cotwo\ correlation. 

We find a well-constrained distribution for water from the observed data, $\log \rm H_2O = -3.71^{+0.31}_{-0.24}$, when using the joint 2TP-Crescent retrieval. This is consistent with the 2TP-Fixed model finding within $1\sigma$ but about $3\sigma$ away from the constraint at secondary eclipse. 

\begin{figure*}[!p]
  \centering
  \subfigure[Comparing \water\ constraints: joint vs. \citet{Kreidberg2014} (K14) transmission, emission, and joint \label{fig:label:h2o_a}]{\includegraphics[width=0.46\textwidth]{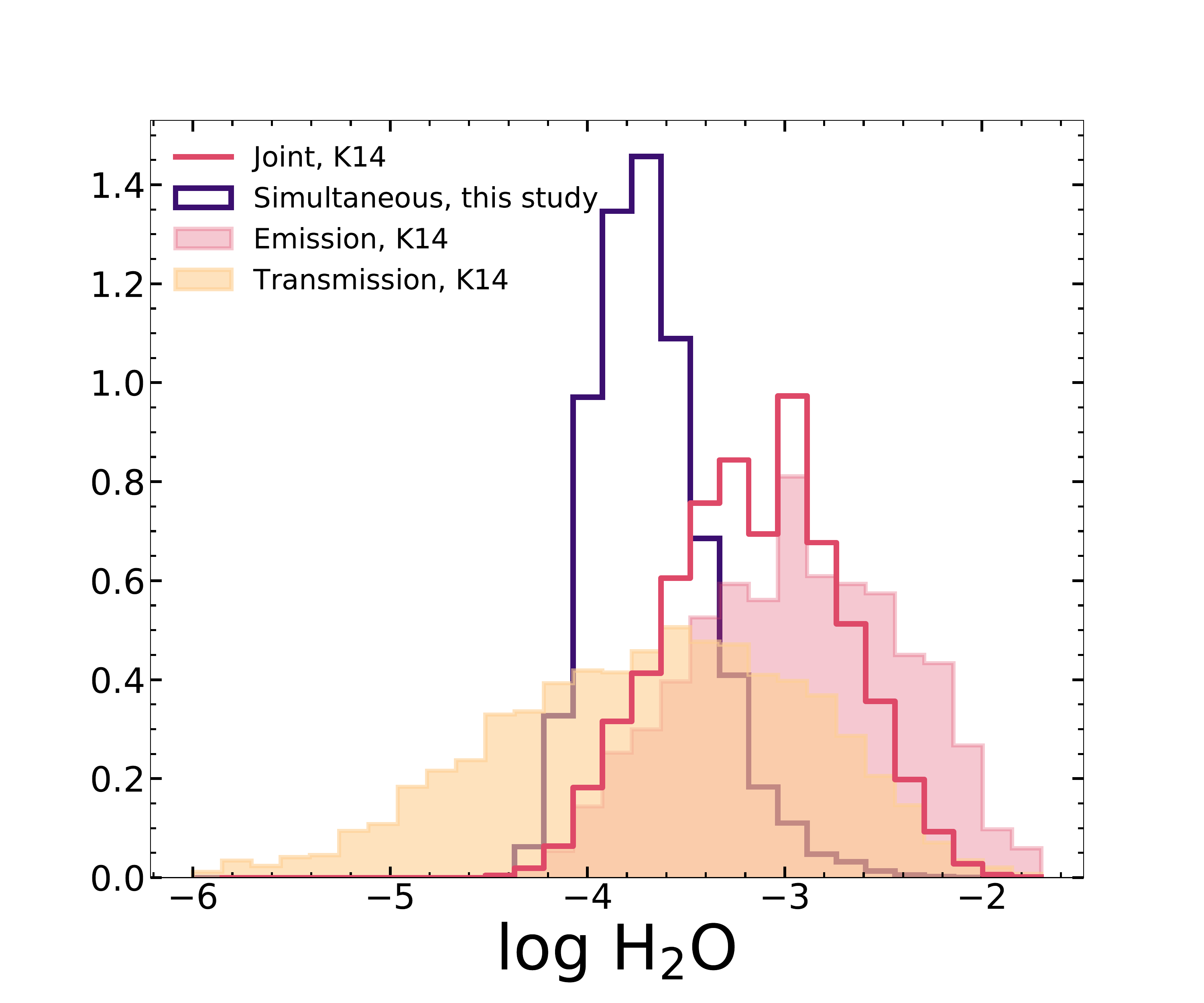}}
  \subfigure[Comparing 1$\sigma$ for \water\ estimates \label{fig:label:h2o_b}]{\includegraphics[width=0.46\textwidth]{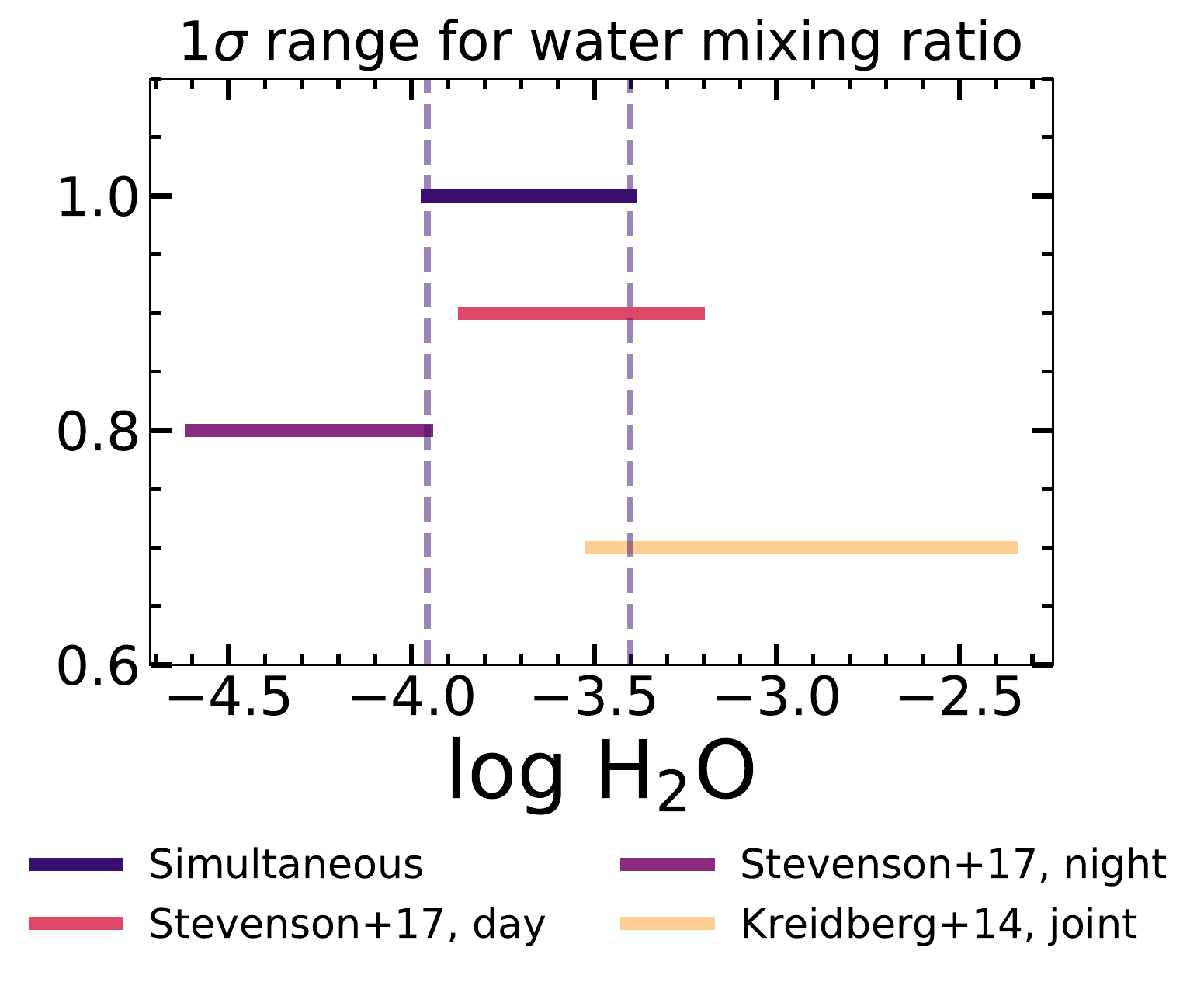}}\hfill
  \caption{Comparing existing \water\ estimates and the estimate from this study using the joint retrieval for WASP-43b. Fig. \ref{fig:label:h2o_a}: We plot the posterior distribution for \water\ from the joint retrieval along with the distributions from \citet{Kreidberg2014}. These include the posterior based on secondary eclipse only, transmission only, and the joint distribution (multiplication of the two posteriors) from the two sets of observations. Fig. \ref{fig:label:h2o_b}: Illustration of the $1\sigma$ range of \water\ estimates from the \citet{Kreidberg2014} joint distribution, \citet{Stevenson2017}, and this study. The \citet{Stevenson2017} results are based on multiplying the posteriors from phases grouped as day (first to third quarter) and night and determine corresponding joint \water\ posteriors. Vertical dashed lines are placed to guide the eye during comparison. The joint retrieval constraint of \water\ is lower than the \citet{Kreidberg2014} $1\sigma$ range, but it is overall consistent with their joint distribution and with the dayside estimate from \citet{Stevenson2017}. }
  \label{fig:h2o_constraint}
\end{figure*}

\citet{Kreidberg2014} demonstrated the power of combining posteriors from multiple HST WFC3 data sets (secondary eclipse and transit) to precisely estimate the \water\ abundance for WASP-43b. \citet{Stevenson2017} added Spitzer data in the atmospheric analysis of WASP-43 b. Using a 1TP retrieval model, \citet{Stevenson2017} grouped dayside-dominant and nightside-dominant phases together, providing a \water\ estimate for each as there appeared to be lower \water\ on the nightside.  Figure \ref{fig:h2o_constraint} compares the \water\ distribution from the joint retrieval of WASP-43b spectroscopic phase curve data to the estimates from \citet{Kreidberg2014} and \citet{Stevenson2017}. Our approach differs in that we did not combine posteriors from the phase-by-phase retrievals after the fact but instead utilized the sum of log-likelihoods at each phase to derive a ``self-consistent'' posterior. The joint phase retrieval places the mixing ratio of \water\ to be between a $1\sigma$ range of $1.1\times 10^{-4}$ -- $3.9\times 10^{-4}$, more consistent with the \cite{Stevenson2017} estimate using dayside phases ($1\sigma$ of $1.4\times 10^{-4}$ -- $6.1\times 10^{-4}$). We see less variation as a function of phase in \water\ abundance using the 2TP-Crescent model,  thus we did not differentiate between day and night phases. As seen in Figure \ref{fig:label:h2o_a}, our \water\ estimate is not only consistent with \citet{Kreidberg2014} but is also more precise. We note that we agree with the constraint determined by \citet{Irwin2020} as well, where \water\ is $(2 - 10)\times 10^{-4}$ for WASP-43b based on \citet{Stevenson2017} data.

Another noteworthy result from the joint WASP-43b retrievals is the constraint of \ammonia. Both 2TP-Fixed and 2TP-Crescent approaches agree on $\log \rm NH_3 = -4.89^{+0.29}_{-0.34}$. Strong constraints should always be met with skepticism. However, we note that the retrieved constraint is consistent with expectations from solar-composition disequlibirum chemical models of similar hot Jupiters ({$\sim 1\times10^{-7} - 1\times10^{-5}$ over the atmospheric pressures probed in emission, e.g., \citet{moses2011}}, although this depends on the value of $T_{\rm int}$ \citep{thorngren2019}). 

\begin{figure*}[!tb]
  \centering
  \includegraphics[width=0.9\textwidth]{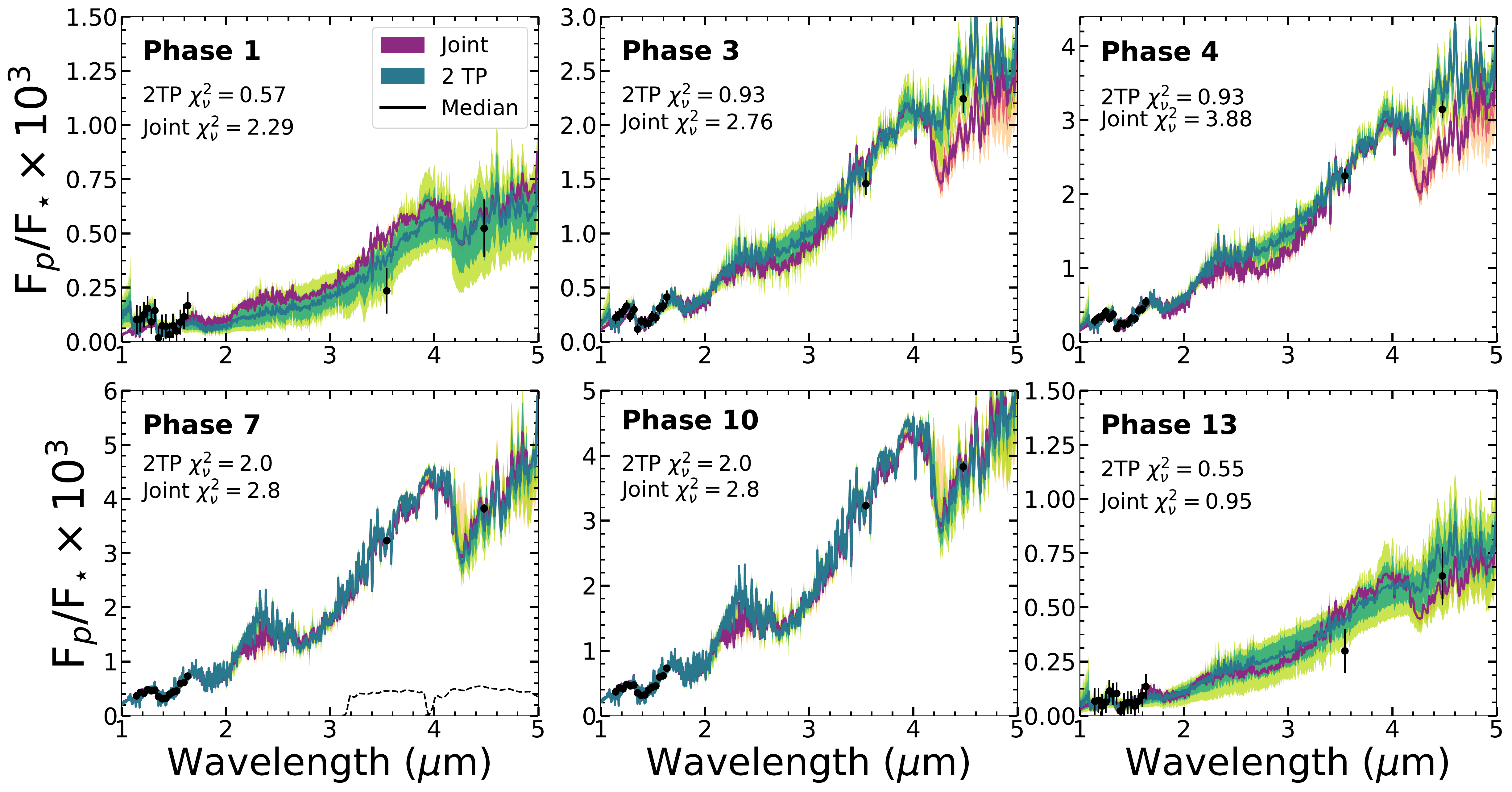}
  \caption{WASP-43b data (HST+Spizter) and high-resolution spectra generated with random posterior draws from the joint retrieval. Shown here are the spectra for phases 1, 3, 4, 7, 10, and 13. Overplotted are the spectral fits from the phase-by-phase 2TP-Crescent retrievals (see Figure \ref{fig:w43_spectra}). We include corresponding $\chi^2_\nu$ values for the Joint and 2TP(-Crescent) cases. In the panel of phase 7, we overplot the Spitzer 3.6$\mu$m and 4.5$\mu$m filter transmission. Jointly-fit spectra are in magenta while 2TP-Crescent spectra are in green. For each set of model spectra, we plot the median, $1\sigma$, and $2\sigma$ contour. Although the constraints are more precise with the joint retrievals, the goodness-of-fit is worse compared to the phase-by-phase scenario. This is expected given only one set of parameters (abundances, TP profiles) were allowed in order to fit all the phases.}
  \label{fig:w43_joint_spectra}
\end{figure*}

Figure \ref{fig:w43_joint_spectra} shows the spectra fits from the joint retrieval of the observed WASP-43b data compared to the 2TP-Crescent phase-by-phase retrievals. The fits from the joint retrieval are poorer fits to the data than the phase-by-phase case. We expect this result because the joint retrieval is restricted by needing to fit the same set of abundances and profiles to different phases at once. The only change from phase to phase in the joint retrieval is the relative area between the hotter and the cooler profile, and thus the shape of the overall spectra from the joint retrieval looks the same at all phases, just at varying levels of total flux. Yet, the use of the full phase curve data set is worth further development since the data from one phase is not independent to the next. Together they paint a holistic image of a planet's atmosphere, and that relationship between phases should be reflected in a retrieval framework for phase curves \citep[e.g.,][]{Irwin2020}. 

\begin{figure*}[!tbp]
  \centering
  \subfigure[TP constraints from joint retrievals, simulated HST/Spitzer]{\includegraphics[width=0.31\textwidth]{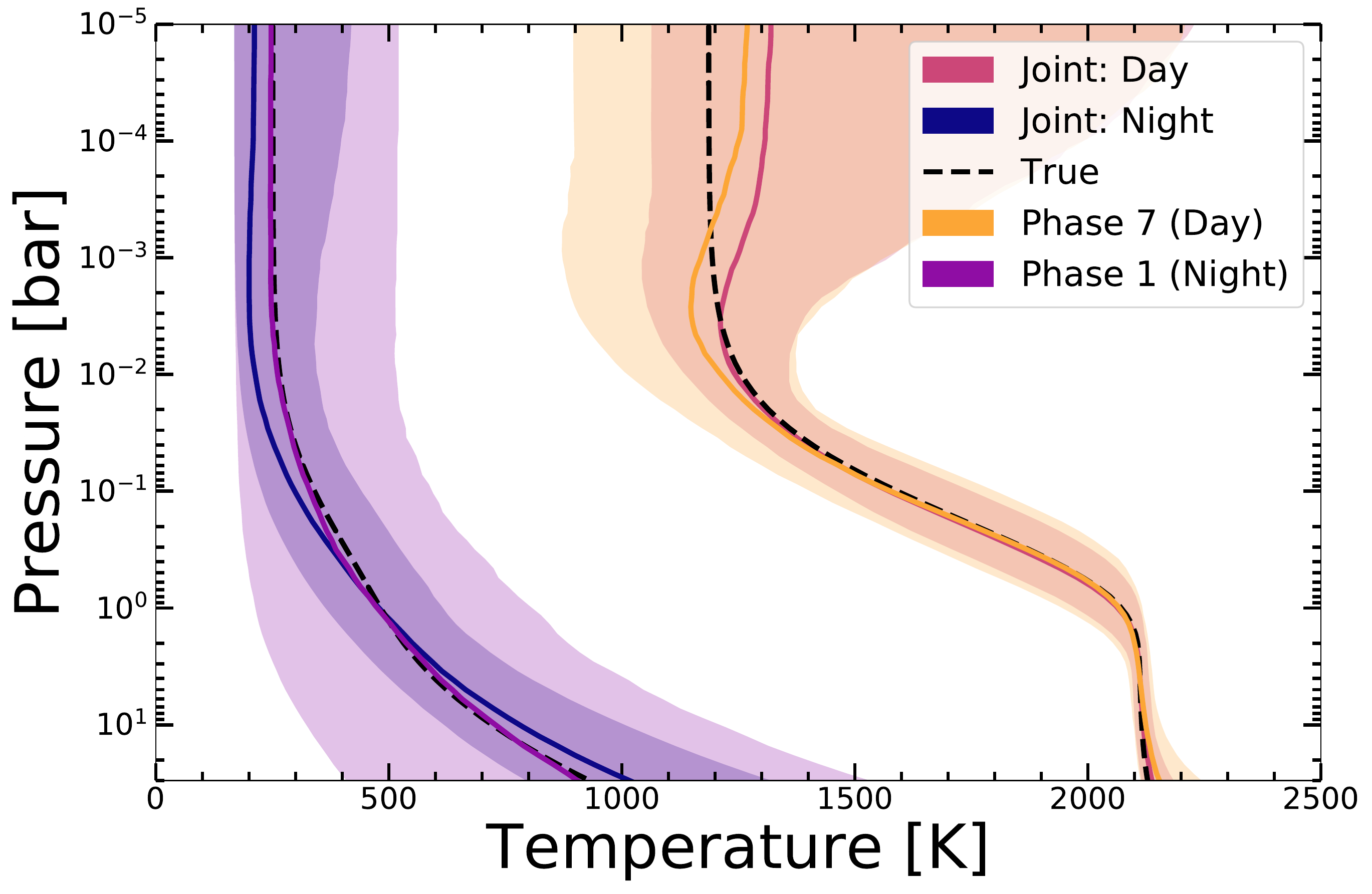}}\hfill
  \subfigure[TP constraints from joint retrievals, WASP-43b]{\includegraphics[width=0.31\textwidth]{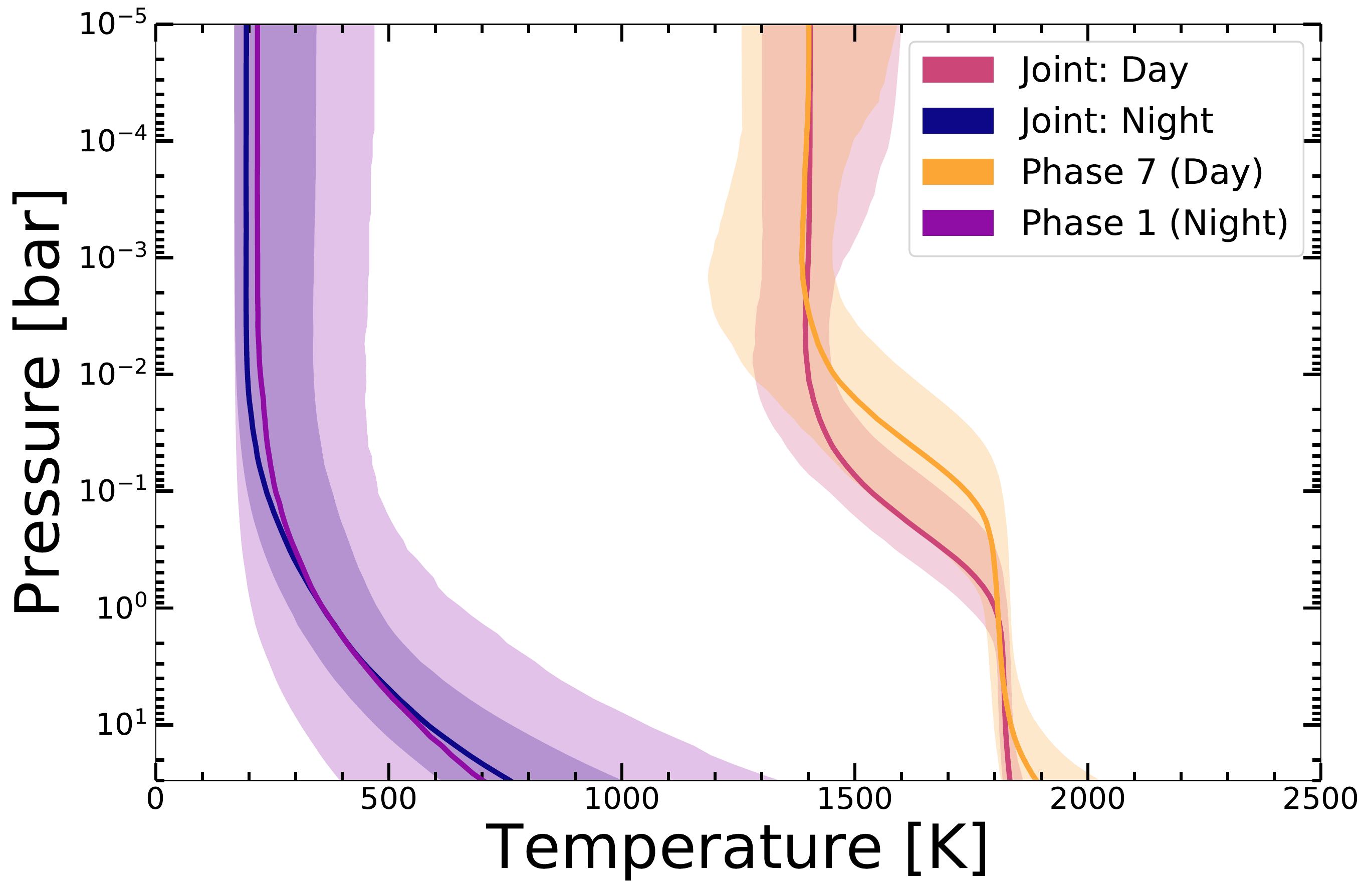}}\hfill
  \subfigure[TP constraints from joint retrievals, simulated JWST]{\includegraphics[width=0.31\textwidth]{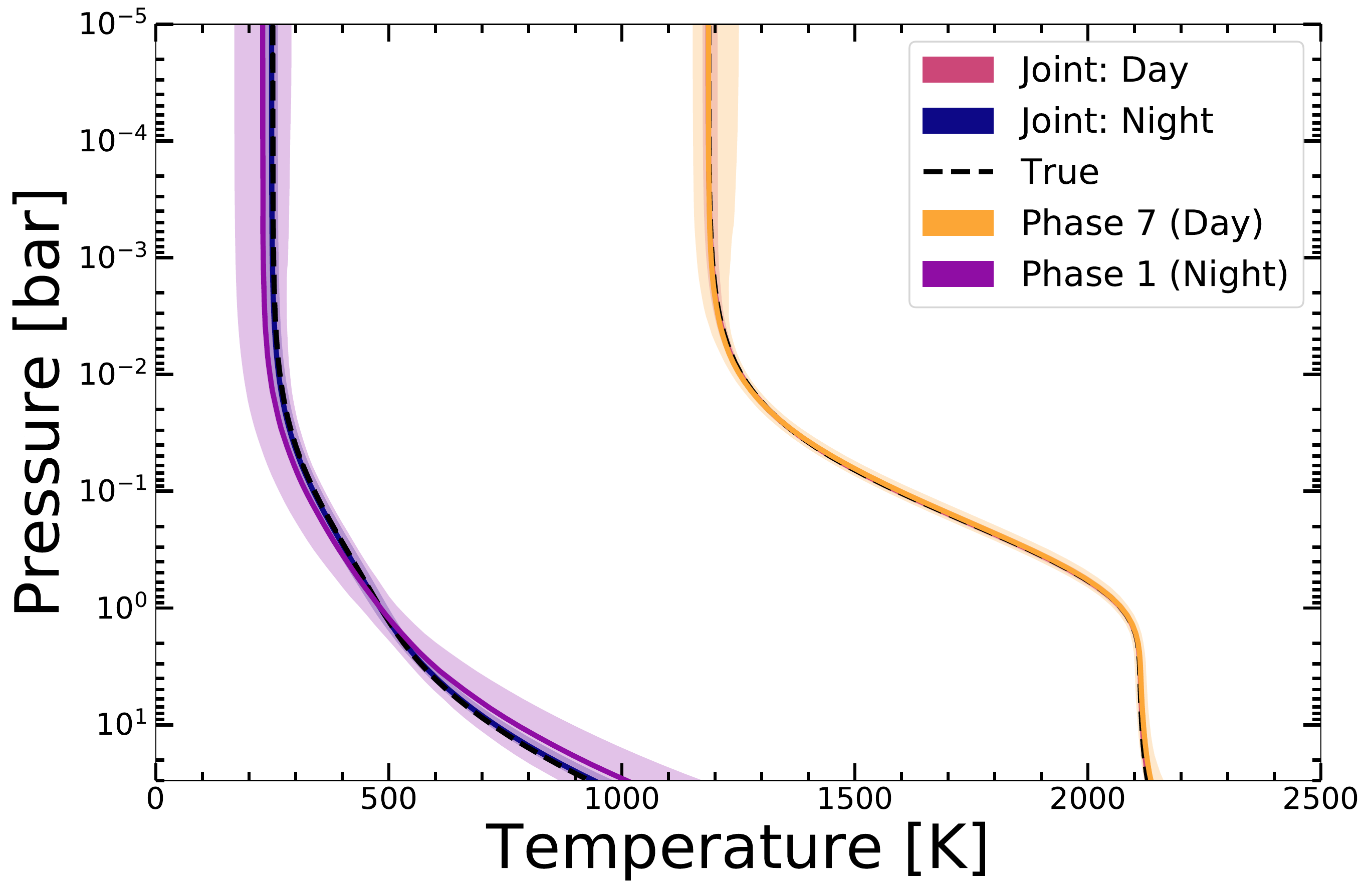}}
  \caption{Constraints of day and night temperature-pressure profiles from joint retrievals of (left) simulated HST/Spitzer data, (middle) observed WASP-43b data, and (right) simulated JWST data. The profiles shown are the median fit and $2\sigma$ envelope of fits from the retrievals. Also included are the phase-by-phase retrieval results of the day profile from phase 7 (secondary eclipse) and night profile from phase 1 (right after transit); see Section \ref{subsec:hst-results}. Dashed lines are the input profiles for the simulated data. The joint retrievals are able to accurately model the true profiles in simulated cases, and provide more precise constraints on the TP profiles than the phase-by-phase retrievals for all data sets.}\hfill
  \label{fig:all_TP}
\end{figure*}

For completeness, in Figure \ref{fig:all_TP}, we show TP profile constraints under the joint retrieval for all three data sets. The joint retrievals place tighter constraints on the TP profiles for all data sets; based on the simulated cases, we can see that the joint retrieval is able to accurately reconstruct the true profiles. That is to say, if an atmosphere is indeed dominated by two contrasting profiles, the joint approach is able to identify them.


Finally, Figure \ref{fig:all_jw} summarizes the abundance constraints resulting from the joint-phase fits for the simulated JWST data. The major advantage seen in simulated retrieval results of JWST data is once again the precision along with the accuracy; here, the precision improvement is almost a factor of two. We also find that 2TP-Fixed and 2TP-Crescent provide similar constraints. This suggests that 2TP-Fixed imposes little bias in the context of our simulated JWST data.

{ Certainly some skepticism is always warranted when combining multiple data sets.  In our case, the most surprising find is the relatively tight constraint on NH$_3$ in the joint-fit compared with the lack of constraint in the phase-by-phase retrievals. One reason for skepticism is the increase in reduced chi-square; increasing by at least 30\% across each phase (Figure \ref{fig:w43_joint_spectra}). Certainly this is to be expected as increased assumptions usually go hand-in-hand with poorer fits.  Similar behaviors are often seen \citep{Wakeford2018, MacMadhu2019} when combining HST STIS and WFC3 data in transit, whereby constraints improve, but fits become worse, with joint constraints in some cases becoming inconsistent with constraints from a single data set alone. Furthermore, we note that there has not yet been a claimed detection of NH$_3$ in the WASP-43b transmission spectra; however, it is worth noting that as the field progresses more and more previously unaccounted for degeneracies readily cloud straight-forward transmission spectral interpretations \citep{lineparmentier2016,caldas2019, WelMadhu2019,MaiLine2019, pluriel2020, macdonald2020, LacyBurrows2020}, presenting a challenge when comparing emission/eclipse constraints to those arising from transmission. }

\section{Discussion and Conclusions}
\label{sec:discussion}
Spectroscopic phase curves offer insight into planetary climate and chemistry by providing a measure of the 2D nature of species abundances and temperature.  We can maximize our leverage of these powerful data sets with atmospheric retrievals.

We have generalized our previous non-homogeneous temperature retrieval methodology \citep{Feng2016} to arbitrary phases using a new geometry scheme. We investigated several TP modeling scenarios: 1TP, 2TP-Crescent, 2TP-Free, and 2TP-Fixed. A 2TP setup uses two profiles to explain the variation in flux as a function of phase while a 1TP setup relies on a changing profile throughout the orbit. We combined these scenarios with different observational setups: simulated HST WFC3 spectroscopy with Spitzer IRAC photometry, actual HST+Spitzer data for WASP-43b from \citet{Stevenson2017}, and simulated JWST data (NIRISS + NIRCam + MIRI LRS). By both using Bayesian model selection and examining the posteriors with respect to the input values for simulated data, we were able to determine which phases need the use of a 2TP model to accurately interpret the atmosphere. 

Our simulated HST+Spitzer data setups provided the following insights:
\begin{itemize}
    \item Even phases closer to transit (i.e., more nightside) can constrain the dayside profile because the hotter profile provides more flux.
    \item We are justified in using a 2TP model for {five out of eight phases (three with moderate evidence, and two with weak)}.
    \item {All} of the phases that favored the 1TP model in model-comparison context returned biased abundance posterior distributions. Specifically, \methane\ appeared artificially constrained at higher values than the input for phases 1 through 5 of the simulated HST+Spitzer data. 
    \item \water\ constraints are robust regardless of model choice. 
    \item Upper limits are placed on CO but {\cotwo\ has biased abundances}; this results from the correlation between the two molecules given overlapping features within the Spitzer photometric bands. 
\end{itemize}

In the case of simulated JWST data:
\begin{itemize}
    \item Every phase strongly justified the use of two profiles except for secondary eclipse (as there is no second TP contribution to the flux). The 1TP model was severely biased in terms of abundance retrievals for all molecules. 
    \item Certain instances of the 1TP model also showed signs of temperature inversion in the profile, adding another layer to its inaccuracy. 
    \item With the 2TP-Crescent model, JWST provides precise constraints on the mixing ratios of \water\ and CO, offering up to a factor of five improvement over HST+Spitzer results. 
    \item We found that the wavelengths of 1.4-3.2 and 3.6-10 micron best differentiated the 1TP and the 2TP models.
\end{itemize}

The important distinguishing wavelength ranges we identified are missing in modern observations. \citet{Taylor2020} examined the information content for different JWST observing modes for a planet with inhomogeneous temperature structure, establishing NIRSpec's observing range ($\sim$1-5 $\mu$m) as especially effective. By considering four different amounts of hot profile contribution, \cite{Taylor2020} also investigated the impact of phase angle, finding prominent abundance biases when relying on the 1D model as we have. When we combine longitudinal information from different phases, we can leverage the retrieval technique to better optimize JWST observations.  

Meanwhile, our application of the different model approaches to the \citet{Stevenson2017} WASP-43b phase curve data reveals:
\begin{itemize}
    \item {Half of the phases have moderate evidence (one of which has weak evidence)} in favor of a 2D model, mostly for phases with more night side visible.
    \item The 2TP-Crescent model retrieved consistent profiles for the day and night sides over the orbit.
    \item The 1TP model finds peaked \methane\ distributions at 8 out of the 15 orbital phases that are reminiscent of the biased distributions seen in the simulated data set.
    \item There is no evidence of CO, while \cotwo\ is constrained. This is likely the same artificial behavior seen in the simulated data caused by the CO-\cotwo\ correlation. 
    \item \water\ is mostly consistent from phase to phase. However, the 2TP-Free model prefers higher values at phases 10-13 than 2TP-Crescent (and 1TP). Thus, the implementation of a more complex model (thermal inhomogeneity in our case) can affect interpretation. 
    \item We identified the evidence of the hot-spot offset on WASP-43b based on the 2TP-Free model results, {pointing to the importance of a flexible geometry in implementing 2D models (i.e., not necessarily pre-determining the day vs. night flux contribution).}

\end{itemize}

We also found that we can combine observed data and our phase-dependent retrieval approach to identify interesting phenomena associated with asymmetry over an orbit. The 2TP-Crescent model retrieves upper limits for \methane\ except at phase 11; with the Free model, that phase is an upper limit instead. The relaxed assumption about day-side emission fraction in the 2TP-Free model allows the model to fit for the slope between the Spitzer points with less day side flux contribution while the 2TP-Crescent model needed significant absorption due to \methane\ to do so. It is helpful to look at the full orbit to spot consistency or outliers. 

Recently, \citet{Mendonca2018} and \citet{morello2019} reanalyzed the Spitzer points from \citet{Stevenson2017} - several of the ``anomalous'' phases (e.g., phase 11, or 0.75) where we detected \methane\ in the 2TP-Crescent model have had their points shifted upwards, particularly the $3.6\, \mu$m band point. While we did not retrieve on this reanalysis to be consistent with the approach in \citet{Irwin2020}, this could yield different results of \methane, CO, and \cotwo\ abundances. Further investigation is warranted; however, this is another reason why retrieving on the full phase curve can be beneficial to provide a more holistic picture of an atmosphere.  

Finally, we introduced the concept of a joint phase curve retrieval and applied that to simulated HST+Spitzer data, the \citet{Stevenson2017} WASP-43b data set, and simulated JWST observations and found:
\begin{itemize}
    \item The 2TP-Crescent and 2TP-Fixed models were consistent in performance, although the 2TP-Fixed model was more prone to biased detection.
    \item \ammonia\ is tightly constrained (about half a dex) to $\sim 10^{-5}$ for WASP-43b by the 2D models. This is an interesting constraint that is plausibly consistent with expectations from disequilibrium chemistry models of similar temperature hot Jupiters. JWST will show if this is real or another bias. 
    \item {Under the assumption of uniform-with-longitude \water\ abundance,} we can place a constraint on \water\ for WASP-43b at $1\sigma$ range of $1.1\times 10^{-4}$ -- $3.9\times 10^{-4}$, increasing precision while remaining consistent with previous studies.
    \item For simulated JWST data, the increase in precision of the constraints compared to the phase-by-phase approach is notable for \water\ and CO, approximately by a factor of two. 
\end{itemize}

Based on these results, it would be best to combine both the phase-by-phase and joint retrieval approaches. Phase-by-phase retrievals identify outliers, providing more insight to accuracy, while the joint can improve precision for molecules with accurate inference. We should also be strategic about applying different models to different phases in order to probe the different temperature structures (e.g., 1D for secondary eclipse). 

\subsection{Future Work}
The value of phase curve retrievals is clear. The promise of richer data sets will necessarily demand the advancement of 3D retrieval techniques, in turn improving our understanding of the 3D-nature of planets. In this study, we used our forward model to generate the data. Onwards, we plan to use 3D GCM models \citep[e.g.,][]{Blecic2017,Irwin2020} to provide the spectra that we retrieve to better identify degeneracies and inform retrieval forward model expansions, including: 

1. Non-uniform chemistry. As our model currently assumes the same chemical composition and distribution for the day and night profiles, an important next upgrade should be to allow differences in the mixing ratios. Bayesian model comparison will once again be important in determining if the data quality justifies the inclusion of the extra parameters associated with more complex chemical profiles. Three-dimensional modelling has shown evidence of species transport \citep{drummond2020}, so retrievals should study its detectability and differentiate between transport and biased constraints. {\citet{macdonald2020} demonstrated biased abundance inferences can arrise when assuming uniform composition for transmission spectra, further motivating the need to inspect this assumption for phase curves. Furthermore,  phase curves provide an avenue for testing the coupled GCM-Chemical Kinetics model hypothesis that molecular abundances homogenize with longitude/latitude in WASP-43b-like temperature objects \cite{coopershowman2006,agundez2012}}. Another important consideration is that we are not seeing the constraint capabilities of observations for molecules such as \cotwo, \methane, or \ammonia, which were chosen to be low in abundance in our study. 

2. Various temperature contrasts. We note that unlike in \citet{Feng2016}, where we vary the temperature contrast between the day and night side, we maintained a fixed, large contrast to better isolate differences. This contrast also models the observed WASP-43b data well. In follow-up work, we suggest implementing multiple contrasts to better model other hot Jupiters. It may be worthwhile to include a means of modeling a more gradual temperature change from the day to night side, as done in \citet{Irwin2020}; this could be done in conjunction with abundance variation. 

3. Clouds. Our scenarios remain cloud-free, so the implementation of clouds in our retrieval framework can extend our study of existing and future data sets. \citet{Irwin2020}'s 2.5D retrieval does not assume clouds either, and infer the presence of clouds on the night side of WASP-43b based on the retrieved low temperatures, which are evident in our retrievals as well. \citet{venot2020} perform cloudy and cloud-free retrievals on simulated JWST MIRI phase curves of WASP-43b. This important groundwork finds that the $5-12\, \mu$m range can confirm or rule out the presence of clouds, although this is dependent on cloud bulk properties and composition. With a variety of aerosol implementation available, there is much to be done in cloudy retrievals \citep{barstow2020}. It is, however, worth noting that strong evidence for relatively cloud-free daysides. 

4. More specialized geometry. We can apply our annulus framework to geometries beyond crescent phases. For instance, we can explore hot spots on the dayside prominent near secondary eclipse, complementing future eclipse mapping studies. \citet{ohno2019} presented the impact of planetary obliquity on phase curves, and our retrieval model can be adapted to study the detection of obliquity. {In addition, the work presented by \citet{changeat2020}, which employs a projected disk for phase curve emission rather than spherical trigonometry as in this work, show the potential in incorporating day-night terminator contribution in phase curve studies. We see potential in a comparison study to investigate that contribution, which is not included in our current model. Much like the importance of benchmarking retrieval models against one another \citet{barstow2020compare}, we expect comparison and synergy among the emerging multi-dimensional retrieval frameworks to be important as we usher in the era of JWST.}

\appendix
\section{Incorporating and validating phase geometry} \label{sec:geometry}
Because our goal is perform computationally intensive retrievals on spectroscopic phase curve data, we need the forward model to remain simple enough while capturing the complexity of the geometry. Here we describe the construction of the 2TP-Crescent model geometry. 

We consider the case of an orbit observed at $N$ phases, where phase angle is defined as $\alpha$. {At a given phase, we leverage the fact that radiative transfer and fluxes are calculated with integration by Gaussian quadrature, accounting for intensities emerging at different viewing angles following \citet{Line2013}}. For an approximation of $N_{\rm Gauss}$ points, we have the same number of intensities, each at an angle of $\mu_i$ with a weight of $w_i$. {The upwelling intensity is given by \citep[Eq. 8]{Line2013}:}

\begin{equation}
    I_\lambda = \sum_{z=0}^{N_{\rm lev}} B_\lambda(T_z) e^{-\sum_{j=z}^{N_{\rm lev}} \Delta \tau_{j,\lambda}}\Delta\tau_{z,\lambda},
\end{equation}

\noindent {where $N_{\rm lev}$ represents the number of pressure levels, $B_\lambda(T_z)$ is the Planck function at wavelength $\lambda$ and temperature in the $z$th level, and $\Delta \tau_{z,\lambda}$ is the total absorption optical depth for all gases in the $z$th level.} We integrate these intensities over $2\pi$ with Gaussian quadrature, leading to an annulus at each $\mu_i$, similar to the formalism from \citet{Barman2005}.

Each annulus can be a linear combination of arbitrary TP profiles: $2\pi\sum{\mu_i w_i I}$. In our study, {$I = A_{\rm hot} I_{\rm hot} + (1 - A_{\rm hot}) I_{\rm cold}$}, where {$A_{\rm hot}$} is the fractional area that is emitting with the hotter profile in a 2TP setup. $I_{\rm hot}$ represents the intensities calculated using a profile based on the TP parameters including $\beta_{\rm day}$, while $I_{\rm cold}$ uses the same parameters except $\beta_{\rm night}$. 

We take advantage of a hot Jupiter's (assumed) tidally locked configuration, which leads to a large temperature contrast, as well as our knowledge of the phase geometry, i.e., the amount of a planet that is seen to be illuminated at angle $\alpha$. The equation for illuminated fraction ($k$), {or emitting fraction $f_{\rm day}$ in our study}, as a function of phase angle ($\alpha$) is 
\begin{equation}
    k \equiv f_{\rm day} = \frac{1 - \cos{\alpha}}{2}.
    \label{eq:illum}
\end{equation}

\begin{figure}[!tbp]
  \centering
  \subfigure[Sideview, defining $\phi_i$.]{\includegraphics[width=0.3\textwidth]{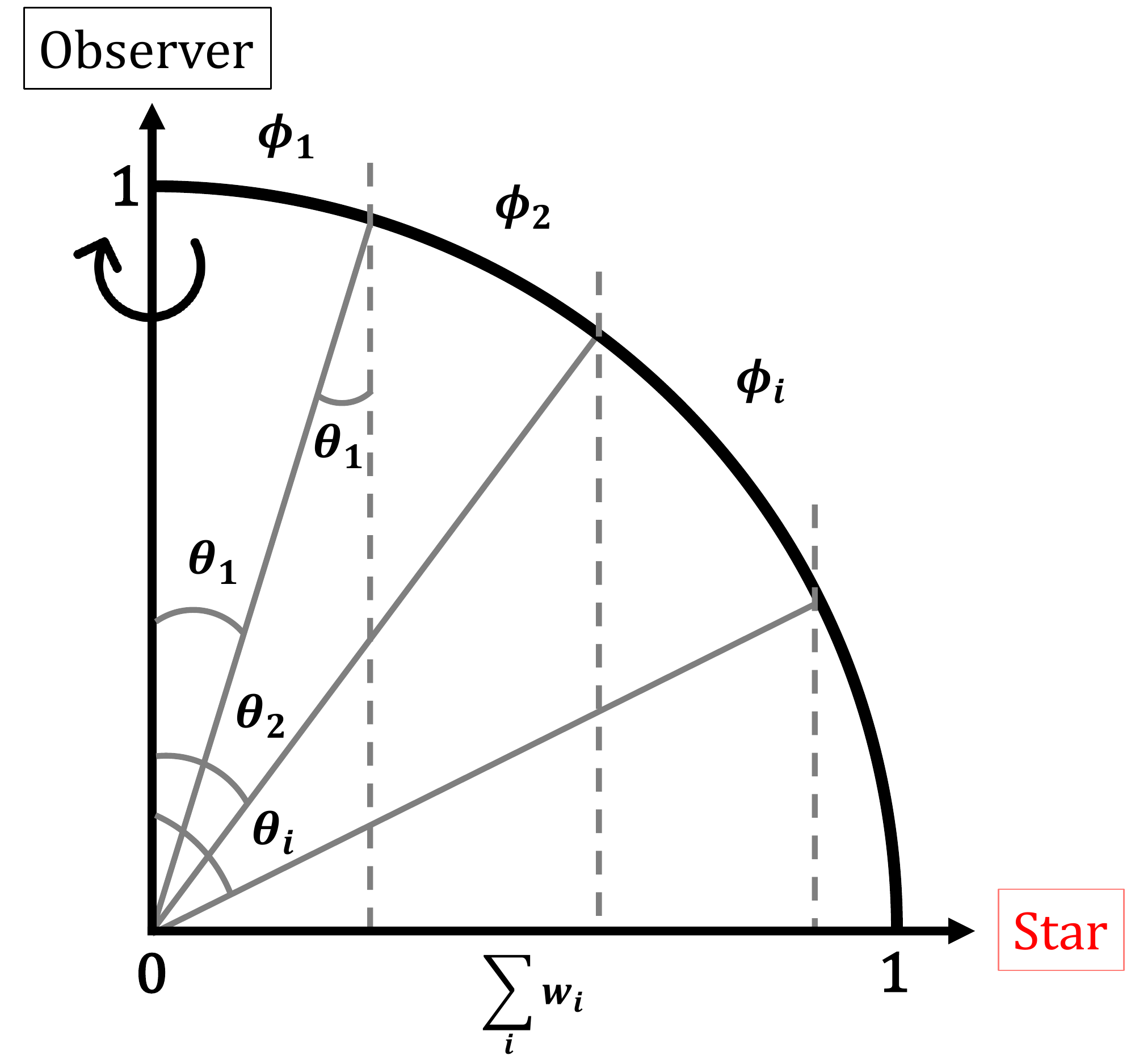}\label{fig:sideview}}\hfill
  \subfigure[Front view, annuli.]{\includegraphics[width=0.3\textwidth]{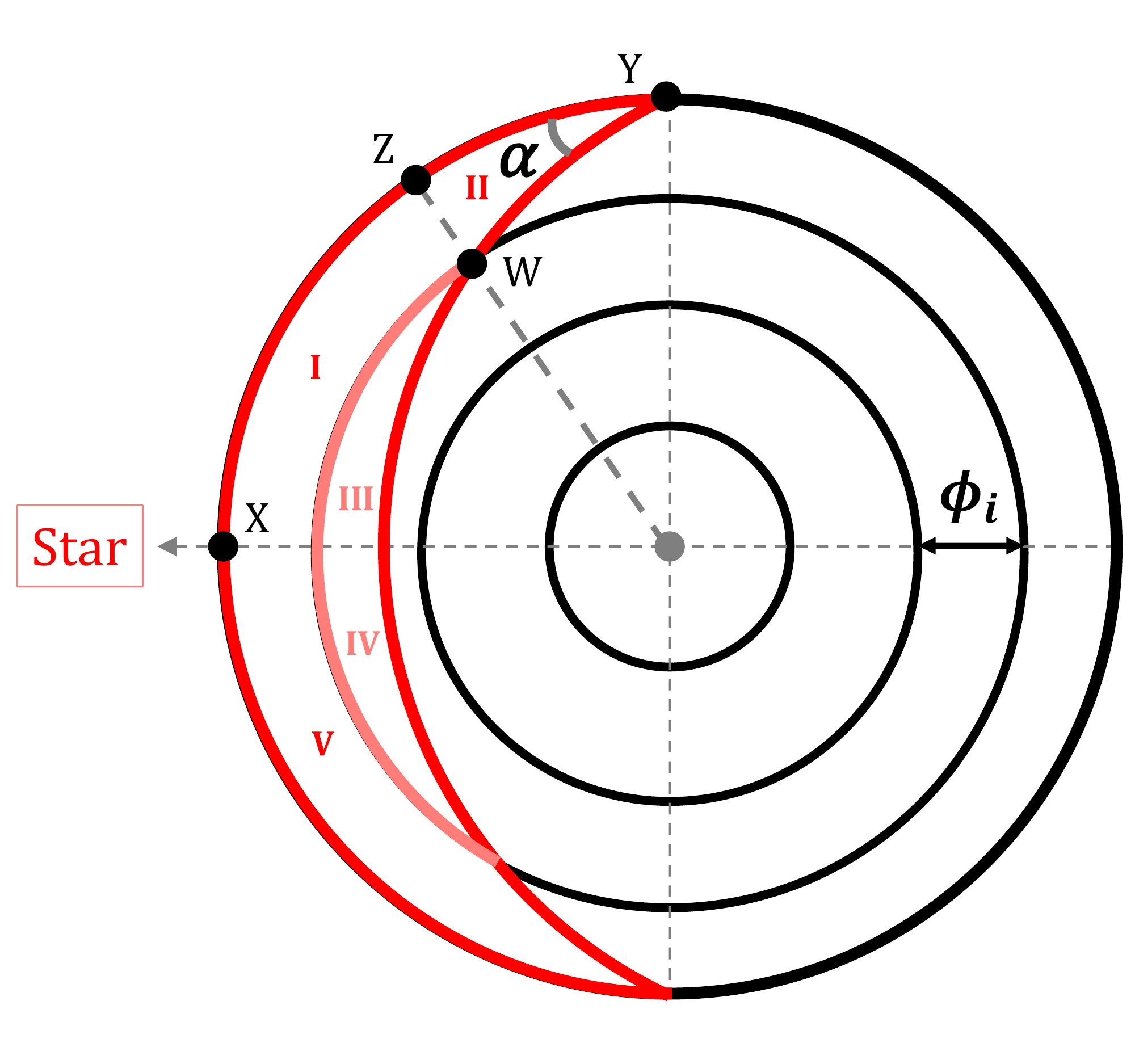}\label{fig:annuli}}\hfill
  \subfigure[AAS spherical triangle solution.]{\includegraphics[width=0.3\textwidth]{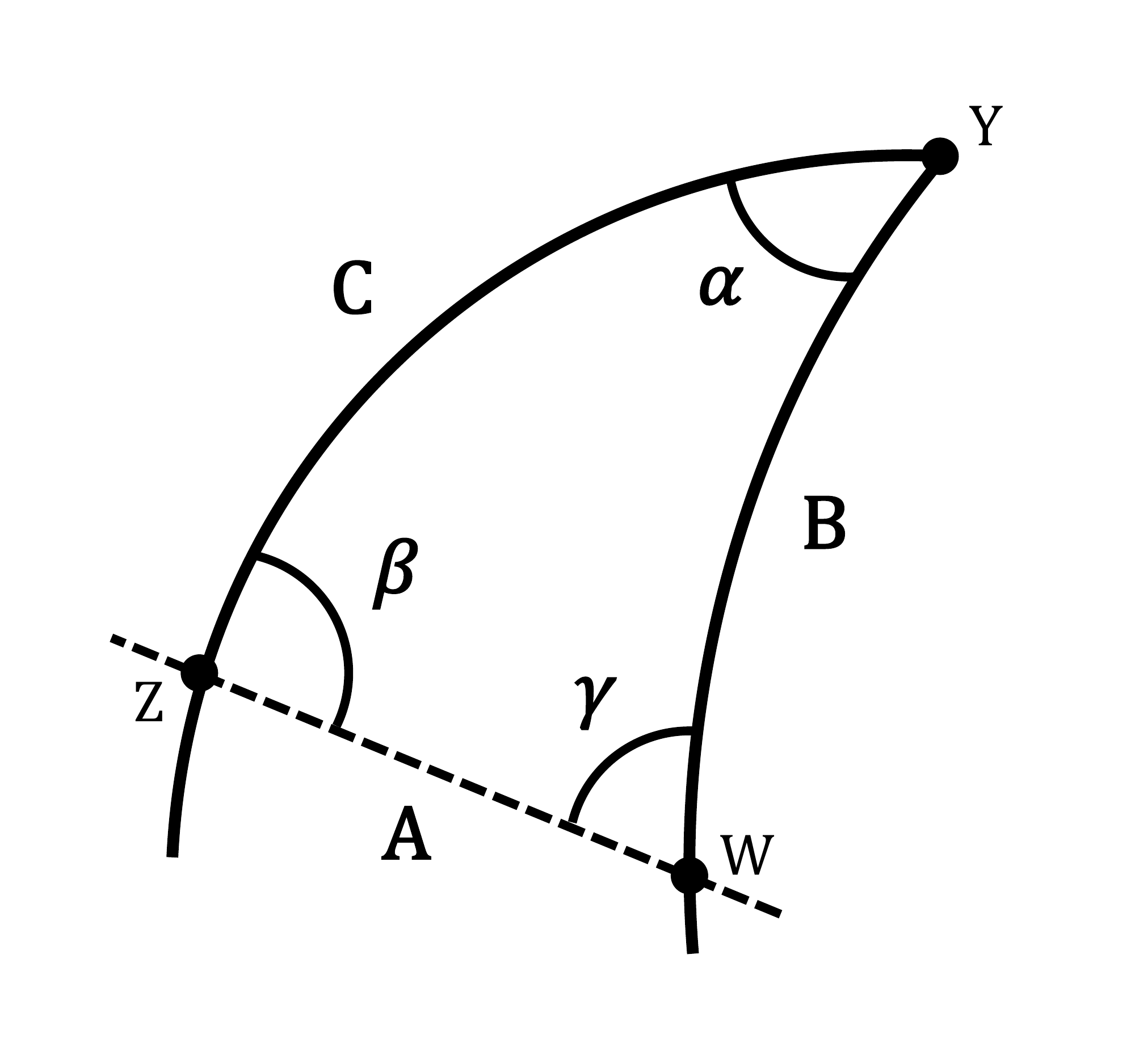}\label{fig:aas}}
  \caption{Detailed schematics of annulus geometry {based on a unit sphere}. (a) 1D sideview of a section of the atmosphere in the Gaussian quadrature setup with $w_i$ for $N$ points. We use a unit circle as an example. For each $\sum_i w_i$, there is a corresponding angle $\theta_i$. The span of each arc between $\theta_i$ is $\phi_i$. Within each arc is a beam of radiation. This quadrant is integrated azimuthally over $2\pi$ to determine the total outgoing radiation of the hemisphere facing the observer. (b) View of hemisphere visible to observer at phase $\alpha$. The emitting region (in red) intersects the annuli at different points. By determining the areas of these segments, we can calculate how much of each annulus is emitting as described in the text. (c) A zoom-in of spherical triangle ZYW for phase $\alpha$. The known variables are $\alpha$, $\beta$, and $A$. This sets up the solution for AAS (angle-angle-side) spherical triangles. See Equations \ref{eq:sideB} to \ref{eq:angleC}.}
  \label{fig:schematic}
\end{figure}

The next step is determining the fraction within each annulus at a given phase that corresponds to the hotter emission on the visible hemisphere. As seen in Figure \ref{fig:sideview} , for the Gaussian quadrature integration of $N$ points, there is a set of weights $w_i$ and corresponding $\theta_i$ that indicates the direction of the radiation beam for each point $i$. {We calculate $\theta_i$ using $\cos{\theta_i} = \frac{\sum_i w_i}{1}$ for a unit sphere. Subsequently, the arc length that spans between two $\theta_i$ is also the angular width of an annulus, or $\phi_i = \theta_i - \theta_{i+1}$, since radius is 1.}

We use the width to determine the area of each annulus, summing up to $2\pi$, the surface area of half a sphere (the visible hemisphere). The covered area of a spherical lune of angle $\alpha$ (at phase $\alpha$) is $2\alpha$ (Fig \ref{fig:annuli}). We can calculate the fractional area in each annulus that then sum up to $2\alpha$ for one profile. 

Figure \ref{fig:annuli} shows a view of $N_{\rm Gauss} = 4$ where we divide the visible hemisphere into annuli, and the planet is viewed at phase $\alpha$. Because the top and bottom halves are symmetric, we only need to determine the illuminated areas of the top and multiply by two. To calculate the area of Region II as marked in Figure \ref{fig:annuli}, we utilize the AAS (angle-angle-side) solution to solving spherical triangles, {assuming a unit sphere}. Spherical AAS is where one Side, one adjacent Angle, and one opposite Angle are known for a triangle on a sphere, as illustrated in Figure \ref{fig:aas}. Because angle $\beta$ is marked by the intersection between a great circle arc from the center of the sphere and the equator (viewed top down), its value is $\beta=\frac{\pi}{2}$. We know $\alpha$ as the phase angle. Side $A$ is the width of the annulus, {or $\phi_{i+1}$ based on Figure \ref{fig:sideview}}. To calculate side $B$, side $C$, and angle $\gamma$, we follow Equations \ref{eq:sideB} through \ref{eq:angleC}. Once side $C$, or arc $ZY$ in Figure \ref{fig:annuli}, is obtained, we get arc $ZX = XY - ZY = \frac{\pi}{2} - C$. {The area of Region II, a spherical triangle, is then $R^2 [(\alpha+\beta+\gamma)-\pi]$, where $R=1$ for our assumed unit sphere.} The area of Region I is calculated as the area of a longitude-latitude patch, using $I = ZX \cdot \left|\sin{\rm lat1} - \sin{\rm lat2}\right|$, where $\rm lat1 = \frac{\pi}{2}$, $\rm lat2 = \theta_i$ for the outermost annulus in Figure \ref{fig:sideview}. The area of Region III is then $\alpha - (I + II)$. Based on these methods, we calculate the set of areas for hot regions in the annuli as a function of phase. 

\begin{equation}
    B = \frac{\sin{A}\sin{\beta}}{\sin{\alpha}}
    \label{eq:sideB}
\end{equation}

\begin{equation}
    C = 2\arctan\Big[\tan{\big(\frac{1}{2}(A-B)\big)}\frac{\sin{\big(\frac{1}{2}(\alpha+\beta)\big)}}{\sin{\big(\frac{1}{2}(\alpha-\beta)\big)}}\Big]
    \label{eq:sideC}
\end{equation}

\begin{equation}
    \gamma = 2\arccot\Big[\tan{\big(\frac{1}{2}(\alpha-\beta)\big)}\frac{\sin{\big(\frac{1}{2}(A+B)\big)}}{\sin{\big(\frac{1}{2}(A-B)\big)}}\Big]
    \label{eq:angleC}
\end{equation}

An important choice in balancing forward modeling speed and model accuracy is setting $N_{\rm Gauss}$. We first verify that our annulus model produces the emitting fractions as determined by Equation \ref{eq:illum} for the set of phase angles in Table \ref{tab:phases}. A 3D model with higher spatial resolution and individually calculated fluxes would better simulate a realistic planet atmosphere at partially illuminated phases. However, 3D models are time-consuming to run even once. If one attempts to fold that into a retrieval framework, where numerous calls to the forward model are necessary, then it will not be an effective way to estimate properties of the atmosphere. However, we can make use of 3D models to validate our annulus approach. If the spectrum from the annulus method matches that of a 3D output, then we would be confident in the retrieval inferences \citep[similar to][]{Blecic2017}. 

We compare the spectra from a 3D model to spectra generated with the annulus model using $N_{\rm Gauss} = 4$. Our 3D model combines the 1D radiative transfer code \texttt{disort} as adapted by \citet{Morley2015} with a 3D longitude-latitude grid. We assign atmospheric properties and a TP profile to each point in the grid and integrate for the emerging flux from the planet. For the visible hemisphere, the grid has 16 longitudes and 32 latitudes. As orbital phase increases from transit to secondary eclipse, points along additional longitudes adopt the day-side temperature profile.

\begin{figure}[!tbp]
  \centering
  \includegraphics[width=0.85\textwidth]{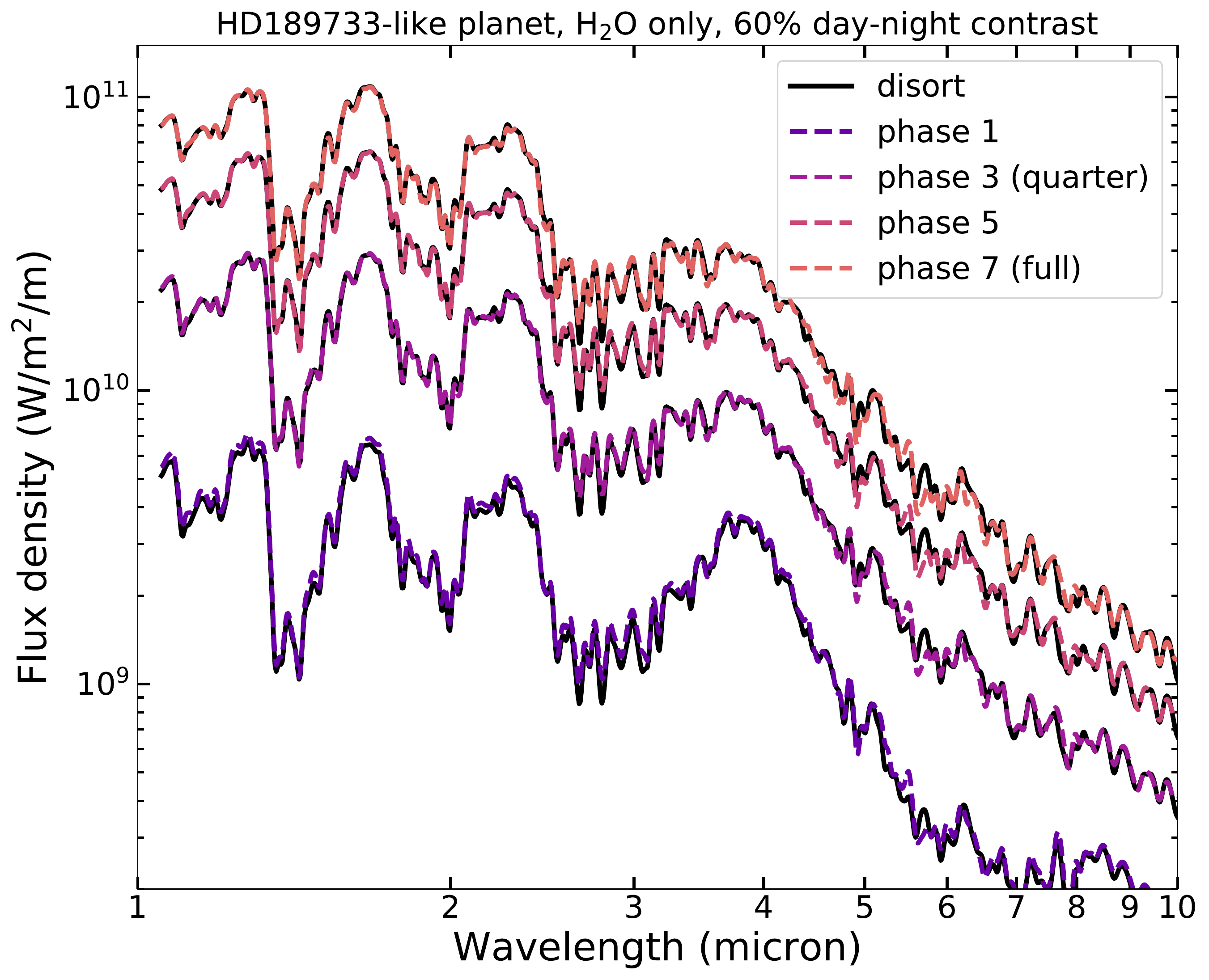}
  \caption{Comparing \citet{Morley2015} {\tt disort} 3D model's spectra {(solid)} to this study's annulus model's spectra {(dashed)} at four phases from after transit to secondary eclipse. The simpler annulus model is able to match the 3D model's output well throughout the orbit.}
  \label{fig:disortVrad2}
\end{figure}

Our test case is an atmosphere with only water vapor as an opacity source with 60\% day-night temperature contrast ($\beta_{\rm night} = 0.4$) for an HD 189733b-like planet. Past phase angle $22.5^{\circ}$ (0.0625 if one phase goes from 0 to 1; see also Table \ref{tab:phases}), we find good agreement between the 3D spectra and our annulus model, as seen in Figure \ref{fig:disortVrad2}. This phase is in fact the smallest we consider in our paper, as it is the first phase in the WASP-43b data we use \citep{Stevenson2017}. We attribute the small mismatches to differences in opacity libraries. Furthermore, our simulated data will be generated and retrieved using the same forward model, providing us with self-consistency in evaluating the results. Thus, $N_{\rm Gauss} = 4$ offers accurate spectra for a given phase geometry and is computationally efficient within our retrieval framework. We thus use $N_{\rm Gauss} = 4$ for 2TP-Crescent and for all models in this work. 

{ We also assume the day/night effective radii are the same as we do not account for subtle effects, negligible in emission, due to day-to-night scale height variations that more strongly influence transmission spectra \citep{caldas2019, pluriel2020, LacyBurrows2020}. }

\section{Supplementary Results}
\label{sec:supp}
{We report in Table \ref{tab:h2ovals} the median and $1-\sigma$ range on $\log$ \water as a function of phase based on the 1TP and 2TP-Crescent models.}

\def\arraystretch{1.5}
\begin{deluxetable}{cll}
\tablecaption{$\log \rm H_2O$ abundance values from 1TP and 2TP-Crescent Retrievals}
\tablewidth{0.45\textwidth}
\tabletypesize{\scriptsize}
\tablehead{Phase & 1TP  & 2TP-Crescent \\
     &   $\log$ \water     & $\log$ \water }
\startdata
\label{tab:h2ovals}
0	&	$-5.96^{+3.54}_{-3.84}$	&	$-4.83^{+3.06}_{-4.24}$	\\
1	&	$-4.23^{+1.06}_{-0.80}$	&	$-2.67^{+1.23}_{-1.30}$	\\
2	&	$-3.74^{+0.95}_{-0.68}$	&	$-3.07^{+1.15}_{-0.77}$	\\
3	&	$-3.06^{+0.55}_{-0.81}$	&	$-3.40^{+0.62}_{-0.44}$	\\
4	&	$-3.13^{+0.53}_{-0.65}$	&	$-3.48^{+0.50}_{-0.38}$	\\
5	&	$-3.57^{+0.56}_{-0.45}$	&	$-3.49^{+0.43}_{-0.34}$	\\
6	&	$-3.51^{+0.37}_{-0.33}$	&	$-3.45^{+0.36}_{-0.33}$	\\
7	&	$-3.34^{+0.29}_{-0.29}$	&	$-3.37^{+0.27}_{-0.27}$	\\
\enddata
\tablecomments{The input value for $\log \rm H_2O = -3.37$. We report the median and $1-\sigma$ range as a function of phase from the 1TP and 2TP-Crescent models.}
\end{deluxetable}

\begin{figure}[!tbp]
  \centering
  \includegraphics[width=0.95\textwidth]{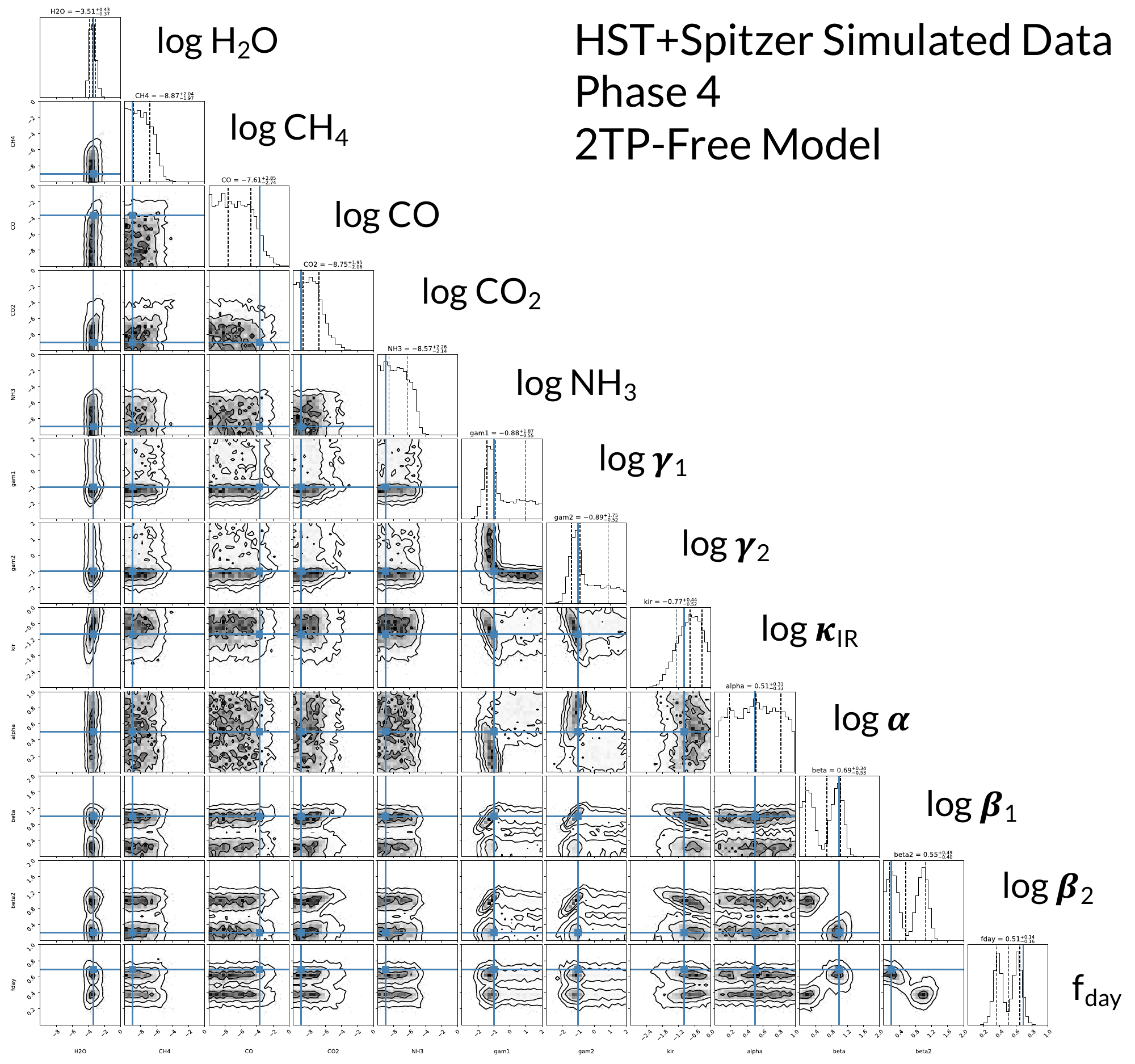}
  \caption{The full retrieval posterior results using the 2TP-Free model for phase 4 of simulated HST+Spitzer data, showing the 1D marginalized posterior distributions and correlations between parameters. Blue solid lines indicate the input value of each parameter. Dashed lines indicate the 16th, 50th, and 84th percentile of each distribution. Each parameter is annotated by the median and $1-\sigma$ (68\%) interval of its posterior.}
  \label{fig:fullpost}
\end{figure}

{Here we also include an example of the full posterior distrbutions associated with one of our retrieval runs. Figure \ref{fig:fullpost} shows the posterior distrbutions and parameter correlations for phase 4 simulated HST+Spitzer data analyzed with the 2TP-Free model.}

\acknowledgments
Y.K.F. is supported by the National Science Foundation Graduate Research Fellowship under Grant DGE1339067. M.R.L. and J.J.F acknowledge support for this work from NASA XRP grants NNX17AB56G and 80NSSC19K0446. {We thank the reviewer for providing a thorough, detailed, and thoughtful referee report that improved our manuscript.} Y.K.F. thanks Natasha Batalha and Jake Taylor for their encouragement and insights during the completion of this work. Y.K.F. also thanks Avi Patel and Nikhil Joshi for their participation in this work during the Science Internship Program at UCSC. The computation for this research was performed by the UCSC Hyades supercomputer, supported by the National Science Foundation (award number AST-1229745) and UCSC. We thank Laura Kreidberg, Kevin Stevenson, and collaborators for providing the WASP-43b data. We thank Tom Greene for helping with estimated \textit{JWST} uncertainties. Based on observations made with the NASA/ESA \textit{Hubble Space Telescope} and the NASA \textit{Spitzer Space Telescope}.

\bibliographystyle{apj}

\end{document}